\pdfoutput=1

\documentclass[12pt,a4paper]{article}

\usepackage{multirow}

\usepackage{ifthen} 
\usepackage{subfig} 
\usepackage{pdflscape}
\newboolean{pdflatex}
\setboolean{pdflatex}{true} 

\newboolean{articletitles}
\setboolean{articletitles}{true} 

\newboolean{uprightparticles}
\setboolean{uprightparticles}{false} 

\usepackage{microtype}
\usepackage{lineno}  
\usepackage{xspace} 

\usepackage{graphicx}  
\usepackage{color}
\usepackage{colortbl}
\graphicspath{{./figs/}} 

\usepackage{amsmath} 
\usepackage{amssymb}
\usepackage{amsfonts}
\usepackage{upgreek} 

\newcommand*\patchAmsMathEnvironmentForLineno[1]{%
\expandafter\let\csname old#1\expandafter\endcsname\csname #1\endcsname
\expandafter\let\csname oldend#1\expandafter\endcsname\csname
end#1\endcsname
 \renewenvironment{#1}%
   {\linenomath\csname old#1\endcsname}%
   {\csname oldend#1\endcsname\endlinenomath}%
}
\newcommand*\patchBothAmsMathEnvironmentsForLineno[1]{%
  \patchAmsMathEnvironmentForLineno{#1}%
  \patchAmsMathEnvironmentForLineno{#1*}%
}
\AtBeginDocument{%
\patchBothAmsMathEnvironmentsForLineno{equation}%
\patchBothAmsMathEnvironmentsForLineno{align}%
\patchBothAmsMathEnvironmentsForLineno{flalign}%
\patchBothAmsMathEnvironmentsForLineno{alignat}%
\patchBothAmsMathEnvironmentsForLineno{gather}%
\patchBothAmsMathEnvironmentsForLineno{multline}%
}

\usepackage{hyperref}    
\usepackage[all]{hypcap} 

\def\lhcb {\mbox{LHCb}\xspace}

\ifthenelse{\boolean{uprightparticles}}%
{

 \def\Pmu         {\ensuremath{\upmu}\xspace}

 \def\Ppi         {\ensuremath{\uppi}\xspace}

 \def\Ppsi        {\ensuremath{\uppsi}\xspace}

 \def\PDelta      {\ensuremath{\Delta}\xspace}                 
 \def\PXi      {\ensuremath{\Xi}\xspace}                 
 \def\PLambda      {\ensuremath{\Lambda}\xspace}                 
 \def\PSigma      {\ensuremath{\Sigma}\xspace}                 
 \def\POmega      {\ensuremath{\Omega}\xspace}                 
 \def\PUpsilon      {\ensuremath{\Upsilon}\xspace}                 
 
 \def\PB      {\ensuremath{\mathrm{B}}\xspace}                 
                  
 \def\PD      {\ensuremath{\mathrm{D}}\xspace}

 \def\PJ      {\ensuremath{\mathrm{J}}\xspace}                 
 \def\PK      {\ensuremath{\mathrm{K}}\xspace}

 \def\Pb      {\ensuremath{\mathrm{b}}\xspace}                 
 \def\Pc      {\ensuremath{\mathrm{c}}\xspace}

 \def\Pi      {\ensuremath{\mathrm{i}}\xspace}

 \def\Ps      {\ensuremath{\mathrm{s}}\xspace}

}
{

 \def\Pmu         {\ensuremath{\mu}\xspace}

 \def\Ppi         {\ensuremath{\pi}\xspace}

 \def\Ppsi        {\ensuremath{\psi}\xspace}                 
                  
 \mathchardef\PDelta="7101
 \mathchardef\PXi="7104
 \mathchardef\PLambda="7103
 \mathchardef\PSigma="7106
 \mathchardef\POmega="710A
 \mathchardef\PUpsilon="7107
                  
 \def\PB      {\ensuremath{B}\xspace}                 
                  
 \def\PD      {\ensuremath{D}\xspace}

 \def\PJ      {\ensuremath{J}\xspace}                 
 \def\PK      {\ensuremath{K}\xspace}

 \def\Pb      {\ensuremath{b}\xspace}                 
 \def\Pc      {\ensuremath{c}\xspace}

 \def\Pi      {\ensuremath{i}\xspace}

 \def\Ps      {\ensuremath{s}\xspace}

}

\makeatletter
\ifcase \@ptsize \relax
  \newcommand{\miniscule}{\@setfontsize\miniscule{4}{5}}
\or
  \newcommand{\miniscule}{\@setfontsize\miniscule{5}{6}}
\or
  \newcommand{\miniscule}{\@setfontsize\miniscule{5}{6}}
\fi
\makeatother

\DeclareRobustCommand{\optbar}[1]{\shortstack{{\miniscule (\rule[.5ex]{1.25em}{.18mm})}
  \\ [-.7ex] $#1$}}

\def\mup        {{\ensuremath{\Pmu^+}}\xspace}
\def\mun        {{\ensuremath{\Pmu^-}}\xspace} 

\def\squark    {{\ensuremath{\Ps}}\xspace}

\def\cquark    {{\ensuremath{\Pc}}\xspace}

\def\bquark    {{\ensuremath{\Pb}}\xspace}

\def\pion   {{\ensuremath{\Ppi}}\xspace}
\def\piz    {{\ensuremath{\pion^0}}\xspace}

\def\pip    {{\ensuremath{\pion^+}}\xspace}
\def\pim    {{\ensuremath{\pion^-}}\xspace}

\def\kaon    {{\ensuremath{\PK}}\xspace}
  \def\Kbar    {{\kern 0.2em\overline{\kern -0.2em \PK}{}}\xspace}

\def\KorKbar    {\kern 0.18em\optbar{\kern -0.18em K}{}\xspace}

\def\Kp      {{\ensuremath{\kaon^+}}\xspace}
\def\Km      {{\ensuremath{\kaon^-}}\xspace}

\def\Kstarz  {{\ensuremath{\kaon^{*0}}}\xspace}

  \def\Dbar    {{\kern 0.2em\overline{\kern -0.2em \PD}{}}\xspace}
\def\D       {{\ensuremath{\PD}}\xspace}

\def\DorDbar    {\kern 0.18em\optbar{\kern -0.18em D}{}\xspace}

\def\Dm      {{\ensuremath{\D^-}}\xspace}

\def\Dsp     {{\ensuremath{\D^+_\squark}}\xspace}
\def\Dsm     {{\ensuremath{\D^-_\squark}}\xspace}

\def\B       {{\ensuremath{\PB}}\xspace}
\def\Bbar    {{\ensuremath{\kern 0.18em\overline{\kern -0.18em \PB}{}}}\xspace}
\def\Bb      {{\ensuremath{\Bbar}}\xspace}
\def\BorBbar    {\kern 0.18em\optbar{\kern -0.18em B}{}\xspace}
\def\Bz      {{\ensuremath{\B^0}}\xspace}
\def\Bzb     {{\ensuremath{\Bbar{}^0}}\xspace}

\def\Bd      {{\ensuremath{\B^0}}\xspace}
\def\Bs      {{\ensuremath{\B^0_\squark}}\xspace}
\def\Bsb     {{\ensuremath{\Bbar{}^0_\squark}}\xspace}

\def\jpsi     {{\ensuremath{{\PJ\mskip -3mu/\mskip -2mu\Ppsi\mskip 2mu}}}\xspace}

  \def\Y#1S{\ensuremath{\PUpsilon{(#1S)}}\xspace}

\def\Lz          {{\ensuremath{\PLambda}}\xspace}
\def\Lbar        {{\ensuremath{\kern 0.1em\overline{\kern -0.1em\PLambda}}}\xspace}
\def\LorLbar    {\kern 0.18em\optbar{\kern -0.18em \PLambda}{}\xspace}

\def\Lb      {{\ensuremath{\Lz^0_\bquark}}\xspace}
\def\Lbbar   {{\ensuremath{\Lbar{}^0_\bquark}}\xspace}

\newcommand{\decay}[2]{\ensuremath{#1\!\to #2}\xspace}         

\def\to                 {\ensuremath{\rightarrow}\xspace}

\def\CP                {{\ensuremath{C\!P}}\xspace}

\def\BdToJPsiKst  {\decay{\Bd}{\jpsi\Kstarz}}

\def\AT#1     {\ensuremath{A_{\mathrm{T}}^{#1}}\xspace}           

\def\C#1      {\ensuremath{\mathcal{C}_{#1}}\xspace}                       
\def\Cp#1     {\ensuremath{\mathcal{C}_{#1}^{'}}\xspace}                    
\def\Ceff#1   {\ensuremath{\mathcal{C}_{#1}^{\mathrm{(eff)}}}\xspace}        
\def\Cpeff#1  {\ensuremath{\mathcal{C}_{#1}^{'\mathrm{(eff)}}}\xspace}       
\def\Ope#1    {\ensuremath{\mathcal{O}_{#1}}\xspace}                       
\def\Opep#1   {\ensuremath{\mathcal{O}_{#1}^{'}}\xspace}                    

\newcommand{\tev}{\ensuremath{\mathrm{\,Te\kern -0.1em V}}\xspace}
\newcommand{\gev}{\ensuremath{\mathrm{\,Ge\kern -0.1em V}}\xspace}
\newcommand{\mev}{\ensuremath{\mathrm{\,Me\kern -0.1em V}}\xspace}
\newcommand{\kev}{\ensuremath{\mathrm{\,ke\kern -0.1em V}}\xspace}
\newcommand{\ev}{\ensuremath{\mathrm{\,e\kern -0.1em V}}\xspace}
\newcommand{\gevc}{\ensuremath{{\mathrm{\,Ge\kern -0.1em V\!/}c}}\xspace}
\newcommand{\mevc}{\ensuremath{{\mathrm{\,Me\kern -0.1em V\!/}c}}\xspace}
\newcommand{\gevcc}{\ensuremath{{\mathrm{\,Ge\kern -0.1em V\!/}c^2}}\xspace}
\newcommand{\gevgevcccc}{\ensuremath{{\mathrm{\,Ge\kern -0.1em V^2\!/}c^4}}\xspace}
\newcommand{\mevcc}{\ensuremath{{\mathrm{\,Me\kern -0.1em V\!/}c^2}}\xspace}

\def\mm   {\ensuremath{\rm \,mm}\xspace}

\def\mum  {\ensuremath{{\,\upmu\rm m}}\xspace}

\def\invfb   {\ensuremath{\mbox{\,fb}^{-1}}\xspace}

\def\ps   {\ensuremath{{\rm \,ps}}\xspace}
\def\fs   {\ensuremath{\rm \,fs}\xspace}

\def\invps{\ensuremath{{\rm \,ps^{-1}}}\xspace}

\newcommand{\chisq}{\ensuremath{\chi^2}\xspace}

\def\gsim{{~\raise.15em\hbox{$>$}\kern-.85em
          \lower.35em\hbox{$\sim$}~}\xspace}
\def\lsim{{~\raise.15em\hbox{$<$}\kern-.85em
          \lower.35em\hbox{$\sim$}~}\xspace}

\def\pt         {\mbox{$p_{\rm T}$}\xspace}

\def\geant      {\mbox{\textsc{Geant4}}\xspace}

\def\pythia     {\mbox{\textsc{Pythia}}\xspace}

\def\tell1  {TELL1\xspace}
\def\ukl1   {UKL1\xspace}

\usepackage{cite} 
\usepackage{mciteplus}

\begin{document}

\renewcommand{\thefootnote}{\fnsymbol{footnote}}
\setcounter{footnote}{1}


\begin{titlepage}
\pagenumbering{roman}

\vspace*{-1.5cm}
\centerline{\large EUROPEAN ORGANIZATION FOR NUCLEAR RESEARCH (CERN)}
\vspace*{1.5cm}
\hspace*{-0.5cm}
\begin{tabular*}{\linewidth}{lc@{\extracolsep{\fill}}r}
\ifthenelse{\boolean{pdflatex}}
{\vspace*{-2.7cm}\mbox{\!\!\!\includegraphics[width=.14\textwidth]{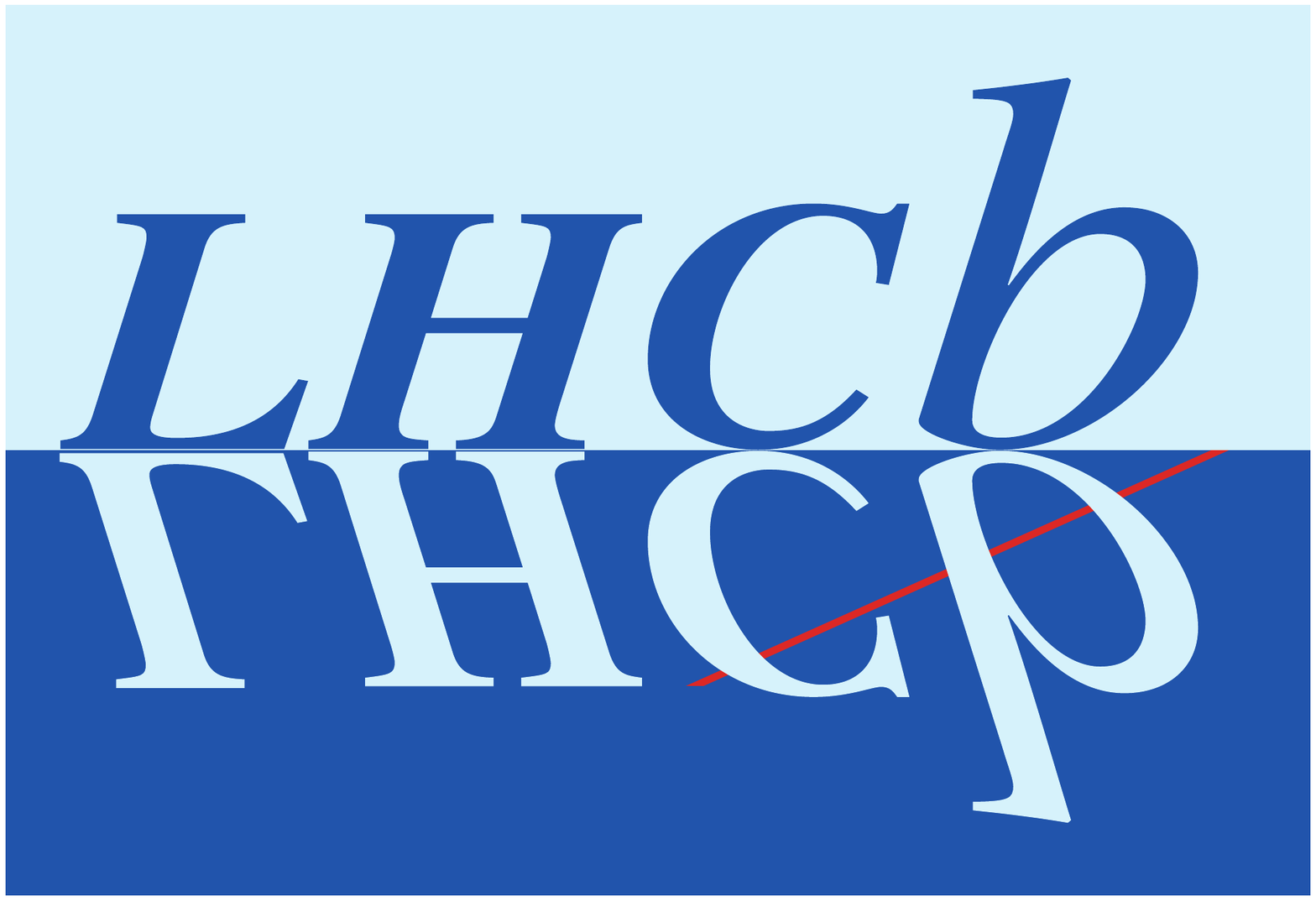}} & &}%
{\vspace*{-1.2cm}\mbox{\!\!\!\includegraphics[width=.12\textwidth]{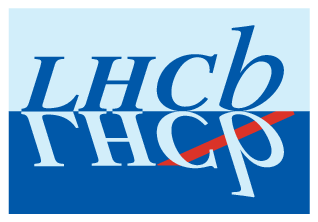}} & &}%
\\
 & & CERN-PH-EP-2014-181 \\  
 & & LHCb-PAPER-2014-042 \\  
 & & July 27, 2014 \\ 
 & & \\
\end{tabular*}

\vspace*{2.0cm}

{\bf\boldmath\huge
  \begin{center}
    Measurement of the $\Bzb-\Bz$ and $\Bsb-\Bs$ production asymmetries in $pp$ collisions at $\sqrt{s}=7\tev$
\end{center}
}

\vspace*{2.0cm}

\begin{center}
The LHCb collaboration\footnote{Authors are listed at the end of this letter.}
\end{center}


\begin{abstract}
  \noindent
The $\Bzb-\Bz$ and $\Bsb-\Bs$ production asymmetries, $A_\mathrm{P}(\Bz)$ and $A_\mathrm{P}(\Bs)$, are
measured by means of a time-dependent analysis of $B^0 \to J/\psi K^{*0}$, $B^0 \to D^- \pi^+$ and $B^0_s 
\to D_s^- \pi^+$ decays, using a data sample corresponding to an integrated luminosity of 1.0\invfb, collected by LHCb in $pp$ collisions at a 
centre-of-mass energy of 7\tev. The measurements are performed as a function of transverse momentum and pseudorapidity of the $B^0$ and $B^0_s$ mesons
within the LHCb acceptance. The production asymmetries, integrated over $p_\mathrm{T}$ and $\eta$ in the range 
$4 < p_\mathrm{T} < 30$\gevc and $2.5<\eta<4.5$,  are determined to be $A_\mathrm{P}(\Bz) = (-0.35 \pm 0.76 \pm 0.28)\%$
 and $A_\mathrm{P}(\Bs) = (1.09 \pm 2.61  \pm 0.66)\%$, where the first uncertainties are statistical and the second systematic.
\end{abstract}

\vspace*{2.0cm}

\begin{center}
  Submitted to Phys. Lett. B
\end{center}

\vspace{\fill}

\vspace{\fill}
 
{\footnotesize
\centerline{\copyright~CERN on behalf of the \lhcb collaboration, license \href{http://creativecommons.org/licenses/by/4.0/}{CC-BY-4.0}.}}
\vspace*{2mm}

\end{titlepage}


\newpage
\setcounter{page}{2}
\mbox{~}

\cleardoublepage


\renewcommand{\thefootnote}{\arabic{footnote}}
\setcounter{footnote}{0}



\pagestyle{plain} 
\setcounter{page}{1}
\pagenumbering{arabic}


%


\section{Introduction}

The production rates of $b$ and $\bar{b}$ hadrons in $pp$ collisions are not expected to be identical. This phenomenon, commonly referred to as the production asymmetry, is related to the fact that there can be coalescence between a perturbatively produced $b$ or $\bar{b}$ quark and the $u$ and $d$ valence quarks in the beam remnant. Therefore, one can expect a slight excess in the production of $B^+$ and $B^0$ mesons with respect to $B^-$ and $\Bzb$ mesons, and \emph{e.g.} of \Lb baryons with respect to \Lbbar baryons. As $b$ and $\bar{b}$ quarks are almost entirely produced in pairs via strong interactions, the existence of $B^+$ and $B^0$ production asymmetries must be compensated by opposite production asymmetries for other $B$-meson and $b$-baryon species. These asymmetries are roughly estimated to be at the 1\% level for $pp$ collisions at LHC energies, and are expected to be enhanced at forward rapidities and small transverse momenta. Other subtle effects of quantum chromodynamics, beyond the coalescence between beauty quarks and light valence quarks, may also contribute~\cite{Chaichian:1993rh,Norrbin:2000jy,Norrbin:2000zc}.

The production asymmetry is one of the key ingredients to perform measurements of $C\!P$ violation in $b$-hadron decays at the LHC, since $C\!P$ asymmetries must be disentangled from other sources. The production asymmetry for $B^0$ and $B^0_s$ mesons is defined as
\begin{equation}
A_\mathrm{P}\left(\B^0_{(s)} \right)\equiv\frac{\sigma\!\left(\Bb^0_{(s)}\right)-\sigma\!\left(B^0_{(s)}\right)}{\sigma\!\left(\Bb^0_{(s)}\right)+\sigma\!\left(B^0_{(s)}\right)},
\end{equation}
where $\sigma$ denotes the production cross-section. Similar asymmetries are also expected when producing charmed hadrons. LHCb has already performed measurements of $D^+-D^-$ and $D^+_s-D^-_s$ production asymmetries, finding values around the 1\% level or less~\cite{LHCb:2012fb,Aaij:2012cy}.

In this paper, the values of $A_\mathrm{P}\left(\B^0 \right)$ and $A_\mathrm{P}\left(\B^0_s \right)$ are constrained by measuring the oscillations of $\Bz$ and $\Bs$ mesons with a time-dependent analysis of the $B^0\rightarrow J/\psi (\mup \mu^- ) K^{*0} (K^+\pi^-)$,  $B^0\rightarrow D^- (\Kp \pim \pim) \pi^+$ and $B_{s}^0\rightarrow D_{s}^- (\Kp K^{-}\pim)\pi^+$ decay rates, without tagging the initial flavour of the decaying $B^0_{(s)}$ meson. The inclusion of charge-conjugate decay modes is implied throughout. The measurements are performed as a function of transverse momentum, $p_\mathrm{T}$, and pseudorapidity, $\eta$, of the $B^0_{(s)}$ meson within the LHCb acceptance, and then integrated over the range $4 < p_\mathrm{T} < 30$\gevc and $2.5< \eta< 4.5$.

\section{Detector, trigger and simulation}
\label{sec:Detector}

The \lhcb detector~\cite{Alves:2008zz} is a single-arm forward spectrometer covering the \mbox{pseudorapidity} range $2<\eta <5$,
designed for the study of particles containing \bquark or \cquark quarks. The detector includes a high-precision tracking system
consisting of a silicon-strip vertex detector surrounding the $pp$ interaction region, a large-area silicon-strip detector located
upstream of a dipole magnet with a bending power of about $4{\rm\,Tm}$, and three stations of silicon-strip detectors and straw drift tubes placed downstream of the magnet.
The tracking system provides a measurement of momentum with a relative uncertainty that varies from 0.4\% at low momentum to 0.6\% at 100\gevc.
The minimum distance of a track to a primary vertex (PV), the impact parameter, is measured with a resolution of $(15+29 /\pt)\mum$, where \pt is in\gevc.
Different types of charged hadrons are distinguished using information from two ring-imaging Cherenkov detectors. 
Photon, electron and hadron candidates are identified by a calorimeter system consisting of scintillating-pad and 
preshower detectors, an electromagnetic calorimeter and a hadronic calorimeter. 
Muons are identified by a system composed of alternating layers of iron and multiwire proportional chambers. The trigger consists of a hardware stage, based on information from the calorimeter and muon systems, followed by a software stage, which applies a full event reconstruction.

In the case of the \BdToJPsiKst decay, events are first selected by a hardware trigger that requires muon candidates with $\pt >1.48$~\gevc. The subsequent software trigger is composed of two stages. The  first stage performs a partial event reconstruction and requires events to have two well identified oppositely charged muons, with invariant mass larger than 2.7\gevcc. The second stage performs a full event reconstruction and only retains events containing a $\mup \mun$  pair that has invariant mass within 120\mevcc of the known $\jpsi$ mass~\cite{PDG2012} and forms a vertex that is significantly displaced from the nearest PV. 

In the case of $B^0 \to D^- \pi^+$ and $B^0_s\to D_s^- \pi^+$ decays, events are first selected by a hardware trigger requiring a high transverse energy cluster in the calorimeter system. 
Events passing the hardware trigger are further filtered by a software trigger which requires a two-, three- or four-track secondary vertex with a large sum of \pt of the tracks and a significant displacement from the PVs. Subsequently, a multivariate algorithm~\cite{BBDT} is applied, aimed at identifying secondary vertices, consistent with the decay of a \bquark hadron.

Simulated events are used to determine the signal selection efficiency, acceptance as function of decay time, decay time resolution, and to model the background.
In the simulation, $pp$ collisions are generated using \pythia 6.4~\cite{Sjostrand:2006za}  with a specific \lhcb configuration~\cite{LHCb-PROC-2010-056}.  
The interaction of the generated particles with the detector, and its
response, are implemented using the \geant toolkit~\cite{Allison:2006ve, *Agostinelli:2002hh} as described in
Ref.~\cite{LHCb-PROC-2011-006}.

\section{Data set and selection}
The selection of \BdToJPsiKst candidates is based on the reconstruction of $\jpsi \to \mup \mun$ and $\Kstarz \to \Kp\pim$ decays. 
The $\jpsi$  candidates are formed from two oppositely charged tracks, identified as muons, having \pt$>$ 500\mevc and originating from a common vertex. The invariant mass of this pair of muons must lie in the range $3030-3150$\mevcc. The $\Kstarz$ candidates are formed from two oppositely charged tracks, one identified as a kaon and the other as a pion, originating from a common vertex. It is required that the $\Kstarz$ candidate has \pt $>$ 1\gevc and that the invariant mass lies in the range $826-966$\mevcc. 

The $\Bz$ candidates are reconstructed from the $\jpsi$  and $\Kstarz$ candidates, with the invariant mass of the $\mup \mun$ pair constrained to the known $\jpsi$  mass. They are required to have an invariant mass in the range $5150-5400$ \mevcc. The decay time of the $\Bz$ candidate is calculated from a vertex and kinematic fit that constrains the candidate to originate from its associated PV~\cite{Hulsbergen:2005pu}.
The $\chisq$ per degree of freedom of the fit is required to be less than 10. Only $\Bz$ candidates with a decay time greater than 0.2\ps are retained.  This lower bound on the decay time rejects a large fraction of the prompt combinatorial background. 

In the case of $B^0 \to D^- \pi^+$ and $B^0_s\to D_s^- \pi^+$ decays, the selection of the $B$-meson candidate is based on the reconstruction of $\Dm \to \Kp\pim\pim$ and $\Dsm \to \Kp \Km\pim$ decays, respectively.  
Requirements are made on the $D^{-}_{(s)}$ decay products before combining them to form a common vertex. The scalar \pt sum of the tracks must exceed 1.8\gevc and the maximal distance of closest approach between all possible pairs of tracks must be less than 0.5\mm. The $D^-_{(s)}$ candidate is required to have a significant flight distance with respect to the associated PV, by requiring a $\chi^{2}$ greater than 36 compared to the zero distance hypothesis. The masses of the $D^-$ and $D_s^-$ candidates must lie within $1850-1890$\mevcc and $1949-1989$\mevcc, respectively. They are subsequently combined with a fourth particle, the bachelor pion, to form the $B$-meson decay vertices. 
The sum of the $D^{-}_{(s)}$ and bachelor pion \pt values must be larger than 5\gevc and the decay time of $B$-meson candidates must be greater than $0.2$ ps.  The cosine of the angle between the $B$-meson candidate  momentum vector and the line segment between the PV and $B$-meson candidate vertex is required to be larger than 0.999.  Particle identification (PID) selection criteria are applied to the kaons and pions from the $D^{-}_{(s)}$  candidate, and to the bachelor pion, in order to reduce the background from other $B$-meson decays with a misidentified kaon or pion and from \Lb decays with a misidentified proton to a negligible level.

A final selection is applied to the candidates that satisfy the criteria described above. It uses a multivariate analysis method~\cite{Breiman,AdaBoost}, optimized separately for each of the three decay modes, to reject the combinatorial background.
The variables used in the selection for the $B$ decay products are  the transverse momentum and the impact parameter. For the $B$ candidates the variables employed are the transverse momentum, the distance of flight and the impact parameter. 

\section{Fit model}
For each signal and background component, the distributions of invariant mass and decay time of $B$-meson candidates are modelled by appropriate probability density functions (PDFs).  
We consider two categories of background: the combinatorial
background, due to the random association of tracks, and the partially reconstructed
background, due to decays with a topology similar to that of the signal, but with one or more particles not reconstructed. The latter is present only for $B ^0_{( s )} \rightarrow D^-_{(s)} \pip$ decays.

\subsection{Mass model}
\label{signalmodel}
The signal component for each decay is modelled convolving a double Gaussian function with a function parameterizing the final state radiation.  
The PDF of the $B$ invariant mass, $m$, is given by
\begin{equation}
g(m)=A\left [\Theta(\mu-m)\,\left (\mu-m \right ) \right]^s \otimes G(m),\label{eq:radcor2}
\end{equation}
where $A$ is a normalization factor, $\Theta$ is the Heaviside function, $G$ is the sum of two Gaussian functions
with different widths and zero mean, and $\mu$ is the $B$ meson mass.
The parameter $s\simeq -0.99$ governs the amount of final state radiation, and is determined using simulated events for each of the three decay modes.
The combinatorial background is modelled by an exponential function for all final states.
In the case of $B^0 \to \Dm \pip$ and $B^0_s \to \Dsm \pip$ decays, a background component due to partially reconstructed
$B^0$ and \Bs decays is also present in the low invariant mass region. The main contributions are expected to come from decays with a missing $\gamma$ or $\pi^0$: $\Bz \rightarrow D^{*-}(\Dm\gamma,\Dm\piz)  \pip$ decays with $\Dm \to \Kp \pim \pim$; $\Bz \rightarrow D^{-} ( \Kp \pim \pim) \rho^{+} (\pip \piz)$ decays; $\Bs \rightarrow D_s^{*-}(\Dsm\gamma,\Dsm\piz) \pip$ decays with $\Dsm \to \Kp \Km \pim$;  $\Bs \rightarrow D_s^{-} (\Kp \Km \pim) \rho^{+} (\pip \piz)$ decays.

We parameterize the partially reconstructed components by means of a kernel estimation
technique~\cite{Cranmer:2000du} based on invariant mass distributions obtained from  full simulation, using the same selection as for data. 
In the case of $\Bs\to\Dsm\pip$ decays, there is also a background component due to $\Bz\to\Dsp\pim$ decays. We account for this component in the fits using the same parameterization adopted for the signal.
The  $\Bz\to\Dsp\pim$ yield is fixed using the ratio between hadronization fractions measured by LHCb\cite{Aaij:2013qqa,LHCb-PAPER-2011-018} and the world average of branching fractions~\cite{PDG2012}. 

\subsection{Decay time model}

The time-dependent decay rate of a neutral $\B^0_{(s)}$ or $\Bb^0_{(s)}$ meson to a flavour-specific $f$ or $\bar{f}$ final state is given by the PDF
\begin{eqnarray}
h\left(t,\psi\right) & =  & K \left(1-\psi A_{\CP} \right)\left(1-\psi A_f\right) \label{eq:untaggedasymmetry}\\
&& \left\{ e^{-\Gamma t}  \left[\Lambda_{+}\cosh\left(\frac{\Delta\Gamma t}{2}\right)+\psi \Lambda_{-} \cos\left(\Delta m t\right) \right] \right\} \otimes R\left( t\right)\epsilon \left( t \right),\nonumber
\end{eqnarray}
where $K$ is a normalization factor, $\epsilon\left( t \right)$ is the acceptance as a function of the decay time, $R \left( t \right) $ is the decay time resolution function, $\Delta m \equiv m_{\rm H}-m_{\rm L}$ and $\Delta\Gamma \equiv\Gamma_{\rm L}-\Gamma_{\rm H}$ are the mass and decay-width differences of the $\B^0_{(s)}-\Bb^0_{(s)}$ system mass eigenstates and $\Gamma$ is the average decay width. The subscripts H and L denote the heavy and light eigenstates, respectively. The two observables are the decay time $t$ and the tag of the final state $\psi$, which assumes the values $\psi=1$ if the final state is $f$ and $\psi=-1$ if the final state is the \CP conjugate $\bar{f}$. 
The terms $\Lambda_{+}$ and $\Lambda_{-}$ are defined as
\begin{equation}
\Lambda_{\pm}\equiv\left(1-A_\mathrm{P}\right)\left|\frac{q}{p}\right|^{1-\psi}\pm\left(1+A_\mathrm{P}\right)\left|\frac{q}{p}\right|^{-1-\psi},
\end{equation}
where $p$ and $q$ are complex parameters entering the definition of the two mass eigenstates of the effective Hamiltonian in the $B^0_{(s)}$ system, $p|B^0_{(s)}\rangle \pm q|\Bb^0_{(s)}\rangle$. The symbol $A_\mathrm{P}$ denotes the production asymmetry of the given $B$ meson, and $A_f$ is the detection asymmetry of the final state, defined in terms of the $f$ and $\bar{f}$ detection efficiencies as
\begin{equation}
A_{f}\equiv\frac{\epsilon_{\bar{f}}-\epsilon_f}{\epsilon_{\bar{f}}+\epsilon_f}.
\end{equation}
The direct \CP asymmetry $A_{\CP}$ is defined as 
\begin{equation}
A_{\CP} \equiv \frac{\mathcal{B}\left(\Bb^0_{(s)}\rightarrow \bar{f}\right) - \mathcal{B}\left( B^0_{(s)} \rightarrow f \right) }{\mathcal{B}\left(\Bb^0_{(s)}\rightarrow \bar{f}\right) + \mathcal{B}\left(B^0_{(s)}\rightarrow f \right)}.
\end{equation}

Trigger and event selections lead to distortions in the shapes of the decay time distributions. The signal decay time acceptances are  determined from
simulated events.  For each simulated decay we apply trigger and selection algorithms as in real data. 

Concerning the combinatorial and the partially reconstructed backgrounds, empirical parameterizations of the decay time spectra are determined by studying the low and high invariant mass sidebands from data. Partially reconstructed backgrounds are only present in the case of $\Bz \to \Dm \pip$  and $\Bs \to \Dsm \pip$ decays. In the case of $\Bs\to\Dsm\pip$ decays, the additional background component due to  $\Bz\to\Dsp\pim$ decays is modelled using the same functional form as that of the $\Bs\to\Dsm\pip$ signal, and the value of the production asymmetry is fixed to that obtained from the $\Bz\to\Dm\pip$ fit.

\subsection{Decay time resolution}

The strategy adopted to study the decay time resolution of the detector consists of reconstructing the decay time of fake \B candidates formed from a $\Dm$ decaying to $\Kp \pim \pim$ and a pion track, both coming from the same PV. 
The bachelor pion must be selected without introducing biases on the decay time, hence
only requirements on momentum and transverse momentum are applied, avoiding the use of impact parameter variables.
The decay time distribution of these fake \B candidates yields an estimate of the decay time resolution of a real decay.
In order to validate the method, simulated events are used for both signals and fake \B decays. The resolution is found to be overestimated  
by about 4\fs. This difference is taken into account as a systematic effect.
The simulation also indicates that a dependence of the resolution on the decay time must be considered. Taking this into
account, an average decay time resolution of $49 \pm 8$\fs is estimated. A resolution model, $R(t)$, consisting of a triple Gaussian function with zero mean 
and three different widths, characterized by an average width of 49\fs, is used. The uncertainty of 8\fs on the average width is taken into account as a systematic uncertainty. It is estimated from simulation that the measurement of the decay time is biased by no more than 2\fs, and the effect is accounted for as a systematic uncertainty.

\section{Determination of the production asymmetries}

The production asymmetry for each of the three decay modes is determined by means of a simultaneous fit to the invariant mass and decay time spectra. To account for the dependence of the production asymmetries on the kinematics of the $B^0$ and $B^0_s$ mesons, each data sample must be divided into bins of ($p_\mathrm{T}$, $\eta$), performing the same fit for each bin.

In order to validate the fit model, a series of fits to the distributions of events obtained from fast simulations is used to verify the accuracy of the central values and the reliability of the uncertainties. No evidence of biases on central values nor of uncertainty misestimations is found. Furthermore, a global fit to the total sample of selected events is performed for each of the three decay modes. The mass differences $\Delta m_d$ and $\Delta m_s$, the mixing parameters $|q/p|_{B^0}$ and $|q/p|_{B^0_s}$, the average decay widths $\Gamma_{d}$ and $\Gamma_{s}$, and the width difference $\Delta\Gamma_{s}$ are fixed to the central values of the measurements reported in Table~\ref{tab:input}. The width difference $\Delta\Gamma_{d}$ is fixed to zero.

According to Eq.~\ref{eq:untaggedasymmetry}, for small values of $A_{\CP}$ and $A_f$, to first order the decay rate is only sensitive to the sum of these two quantities. For this reason, we fix $A_{\CP}$ to zero and leave $A_f$ as a free parameter in the fits. It is empirically verified that the choice of different $A_{\CP}$ values, up to the few percent level, leads to negligible variations of $A_\mathrm{P}$, as expected.

\begin{table}[t]
 \caption{\small Values of the various physical inputs used in the fits.}  
 \begin{center}
   \begin{tabular}{ccc}
\hline
Parameter & Value & Reference\\  
\hline
$\Delta m_d\,[\!\invps]$ &  $\phantom{1}0.510\pm 0.004\phantom{0}$ & \cite{PDG2012}\\
$\Delta m_s\,[\!\invps]$ &  $17.768 \pm 0.024\phantom{0}$ & \cite{Aaij:2013mpa}\\
$\Gamma_{d}\,[\!\invps]$ &  $0.6583 \pm 0.0030$ & \cite{PDG2012}\\
$\Gamma_{s}\,[\!\invps]$ & $0.6596 \pm 0.0046$ & \cite{PDG2012}\\
$\Delta\Gamma_{s}\,[\!\invps]$ &  $\phantom{1}0.081 \pm 0.011\phantom{0}$ & \cite{PDG2012}\\
$|q/p|_{ \Bz }$ & $0.9997\pm 0.0013$ & \cite{bib:HFAG}\\
$|q/p|_{\Bs}$ & $1.0003 \pm 0.0030$ & \cite{Aaij:2013gta}\vspace{1mm}\\
\hline
 \end{tabular}\end{center}
 \label{tab:input}
\end{table}

\begin{figure}[tb]
\begin{center}
    \includegraphics[width=0.48\linewidth]{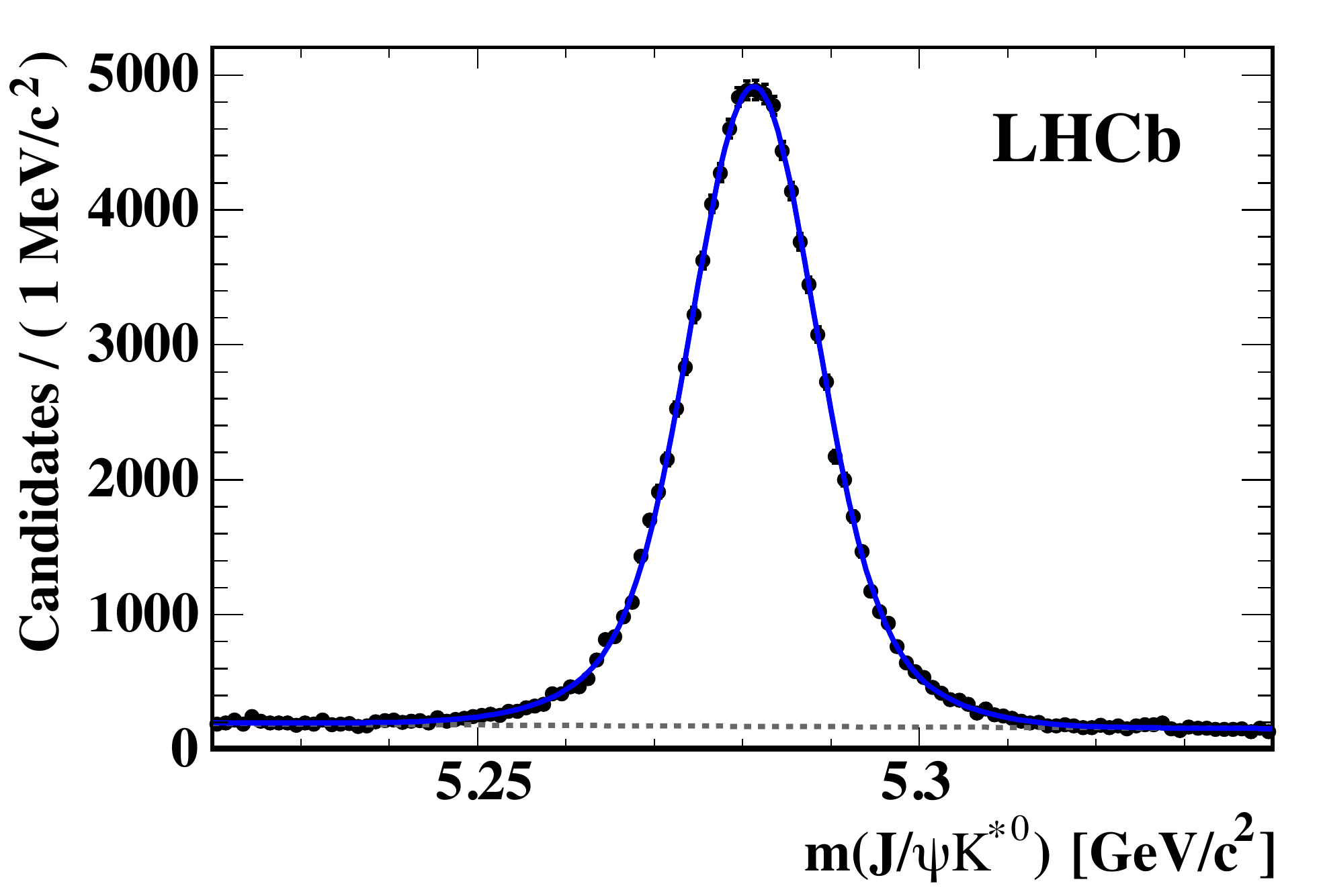}
   \includegraphics[width=0.48\linewidth]{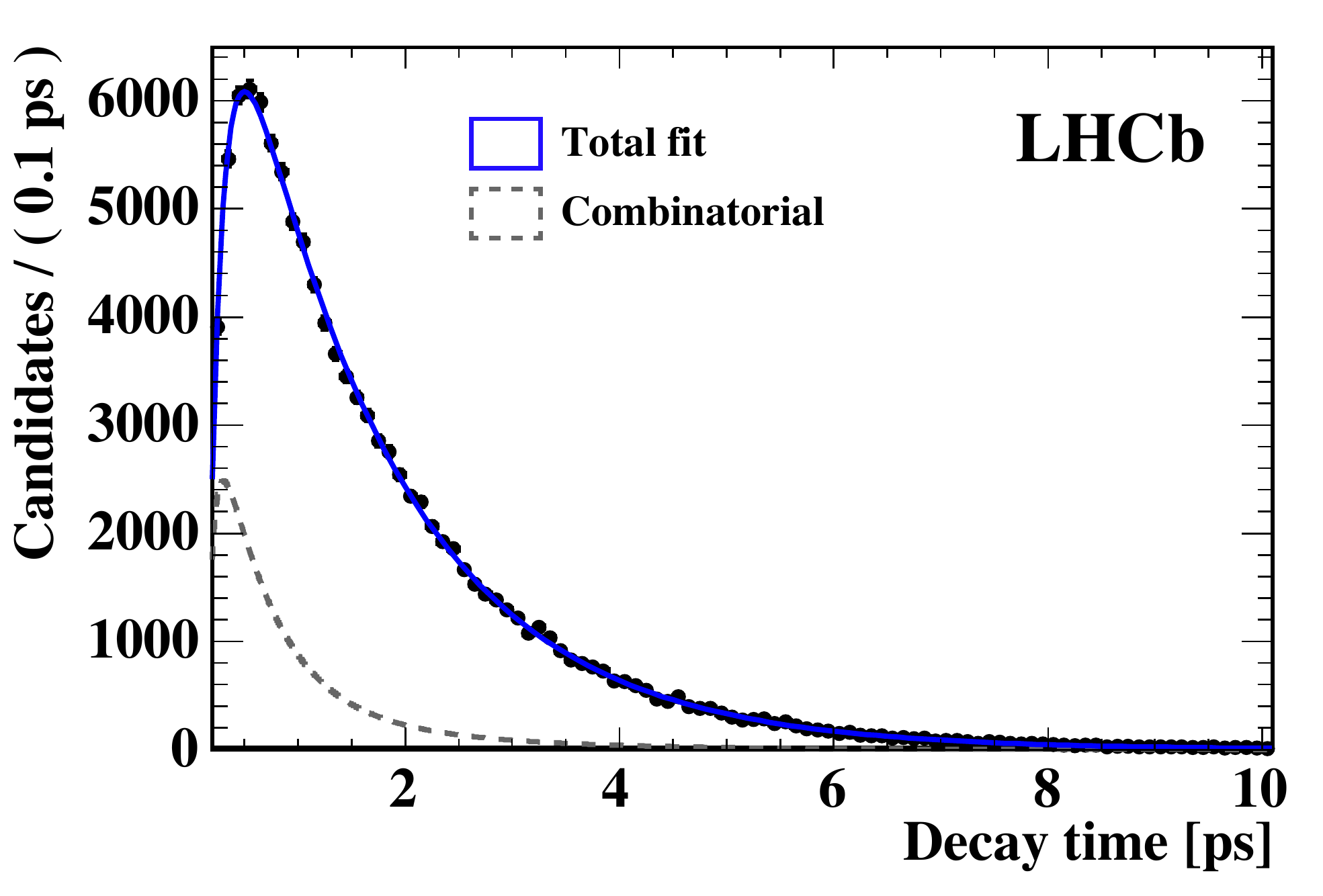}
 \includegraphics[width=0.48\linewidth]{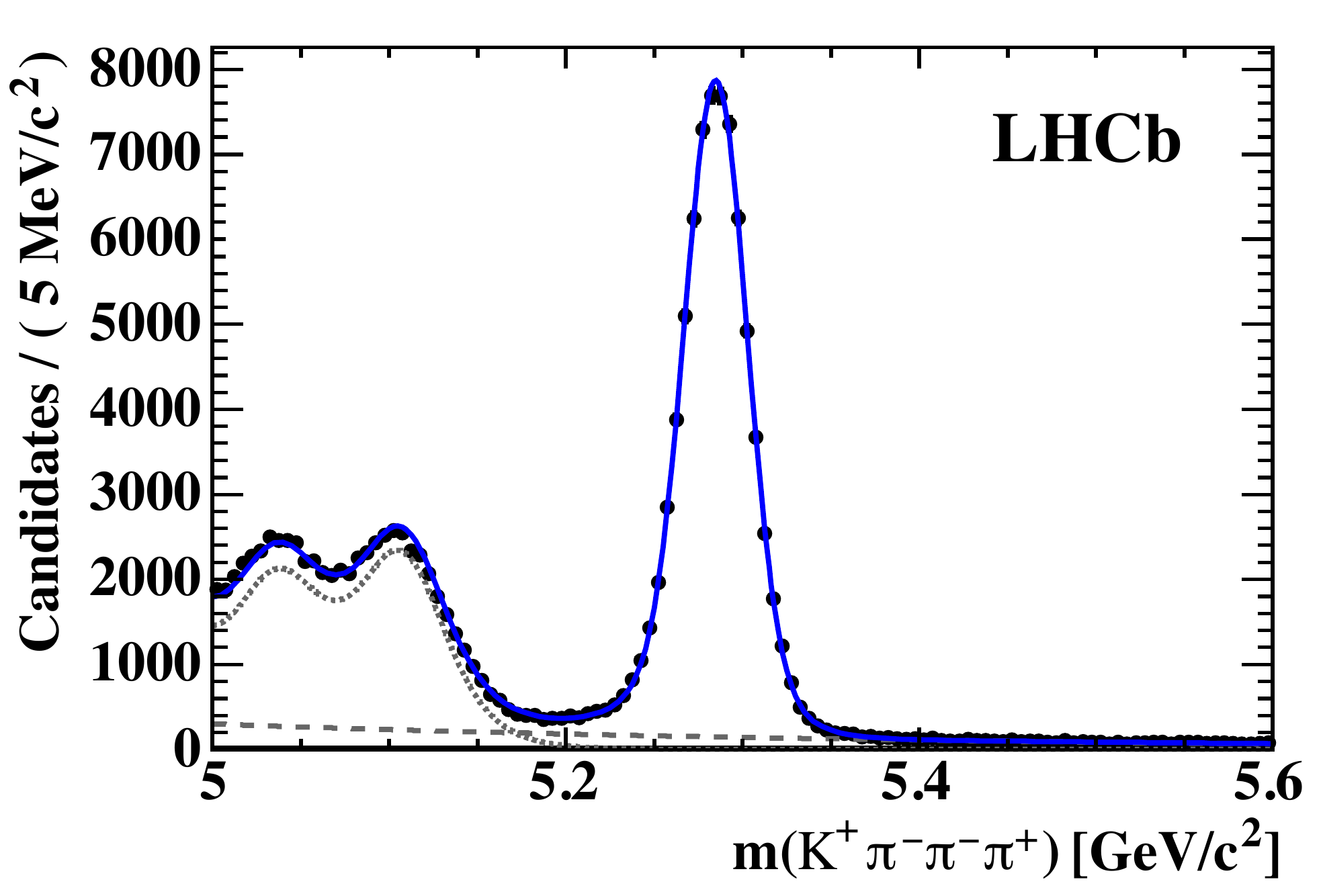}
   \includegraphics[width=0.48\linewidth]{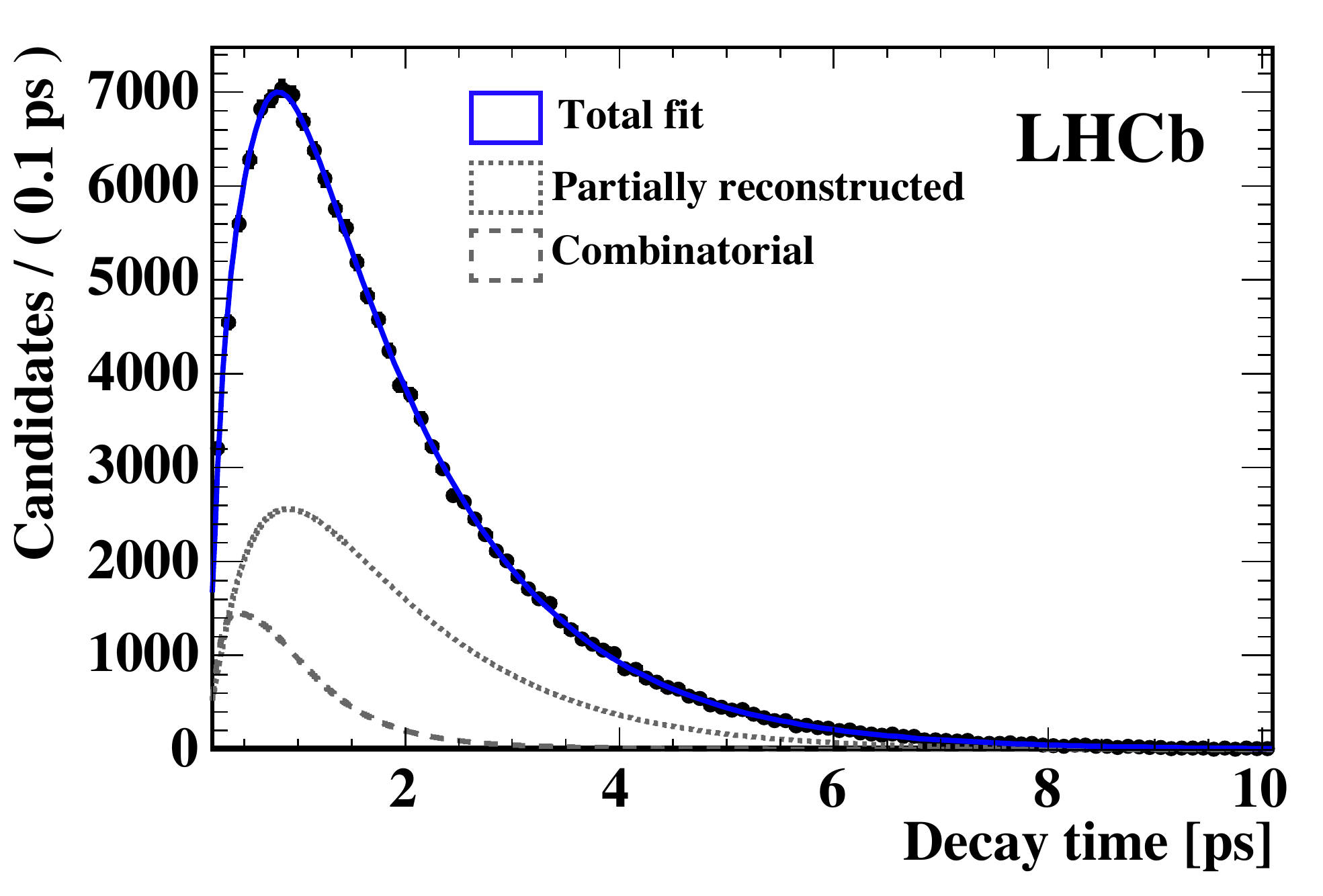}
 \includegraphics[width=0.48\linewidth]{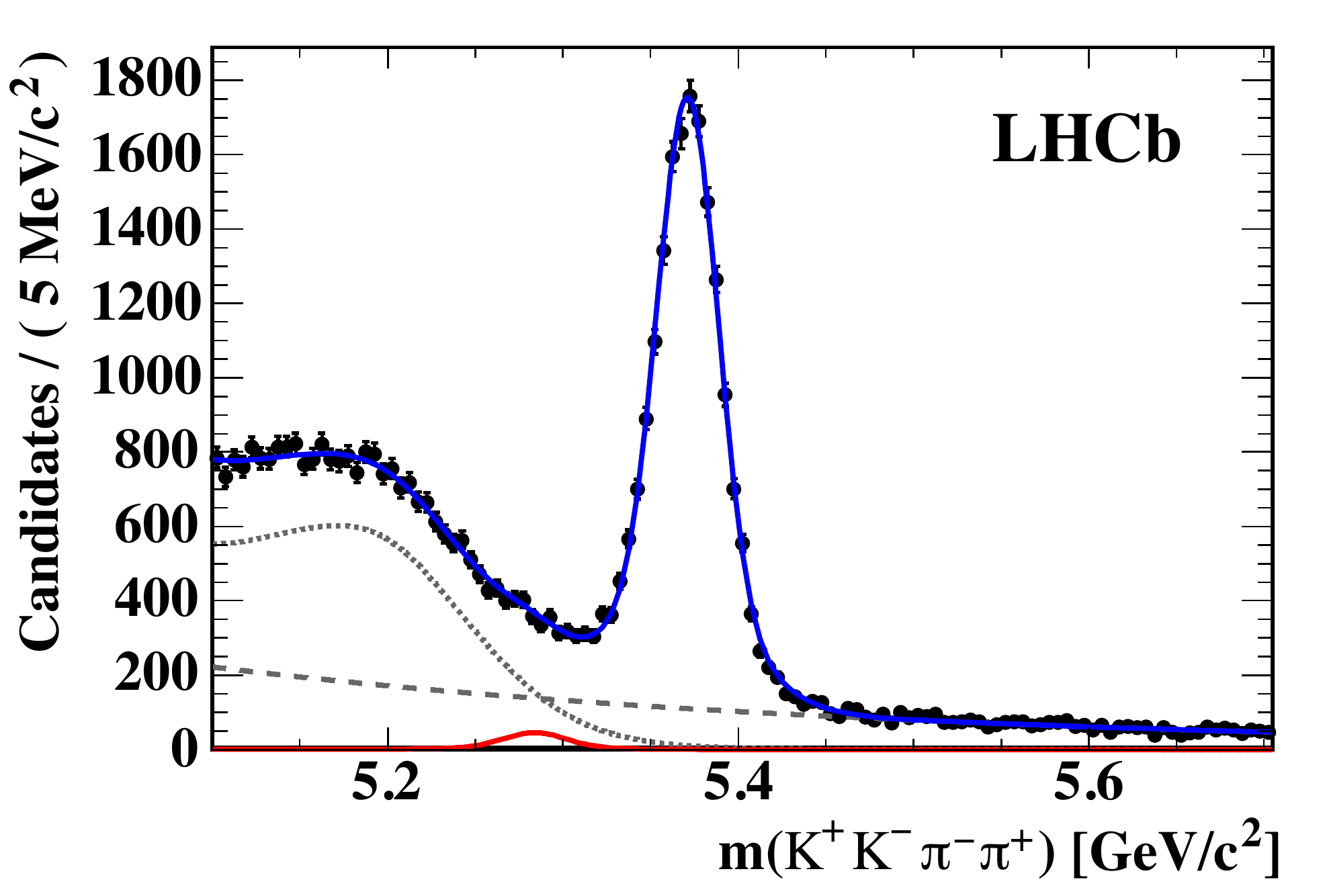}
   \includegraphics[width=0.48\linewidth]{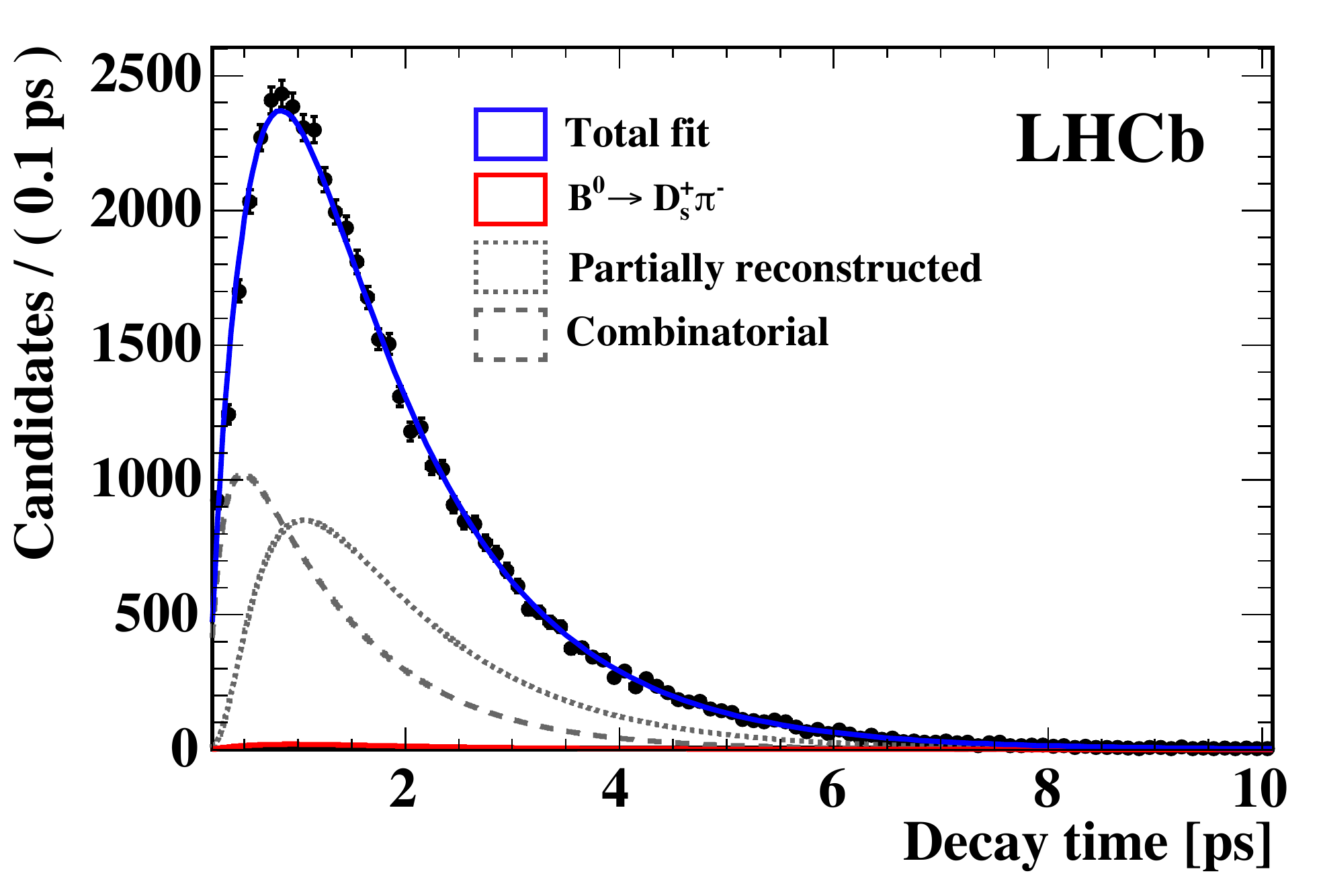}
 \end{center}
  \caption{\small Distributions of (left) invariant mass and (right) decay time for (top) $\Bd \to \jpsi \Kstarz$, (middle) $\Bd \to \Dm \pip$ and  (bottom) $\Bs \to \Dsm \pip$ decays, with the results of the fit overlaid. The contributions of the various background sources are also shown.}  \label{fig:resultBd2JpsiKst}
\end{figure}

\begin{figure}[t]
\begin{center}
    \includegraphics[width=0.48\linewidth]{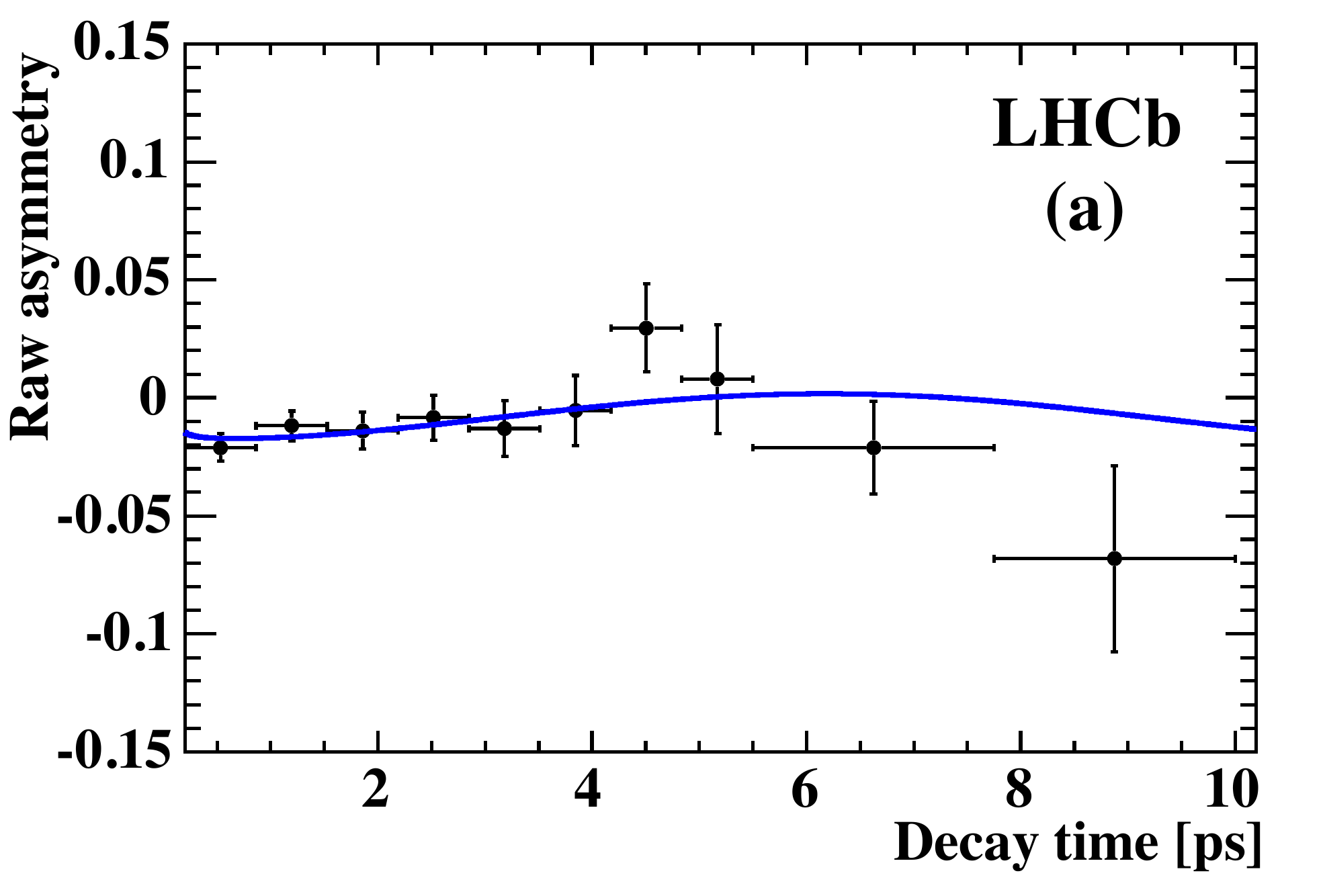}
    \includegraphics[width=0.48\linewidth]{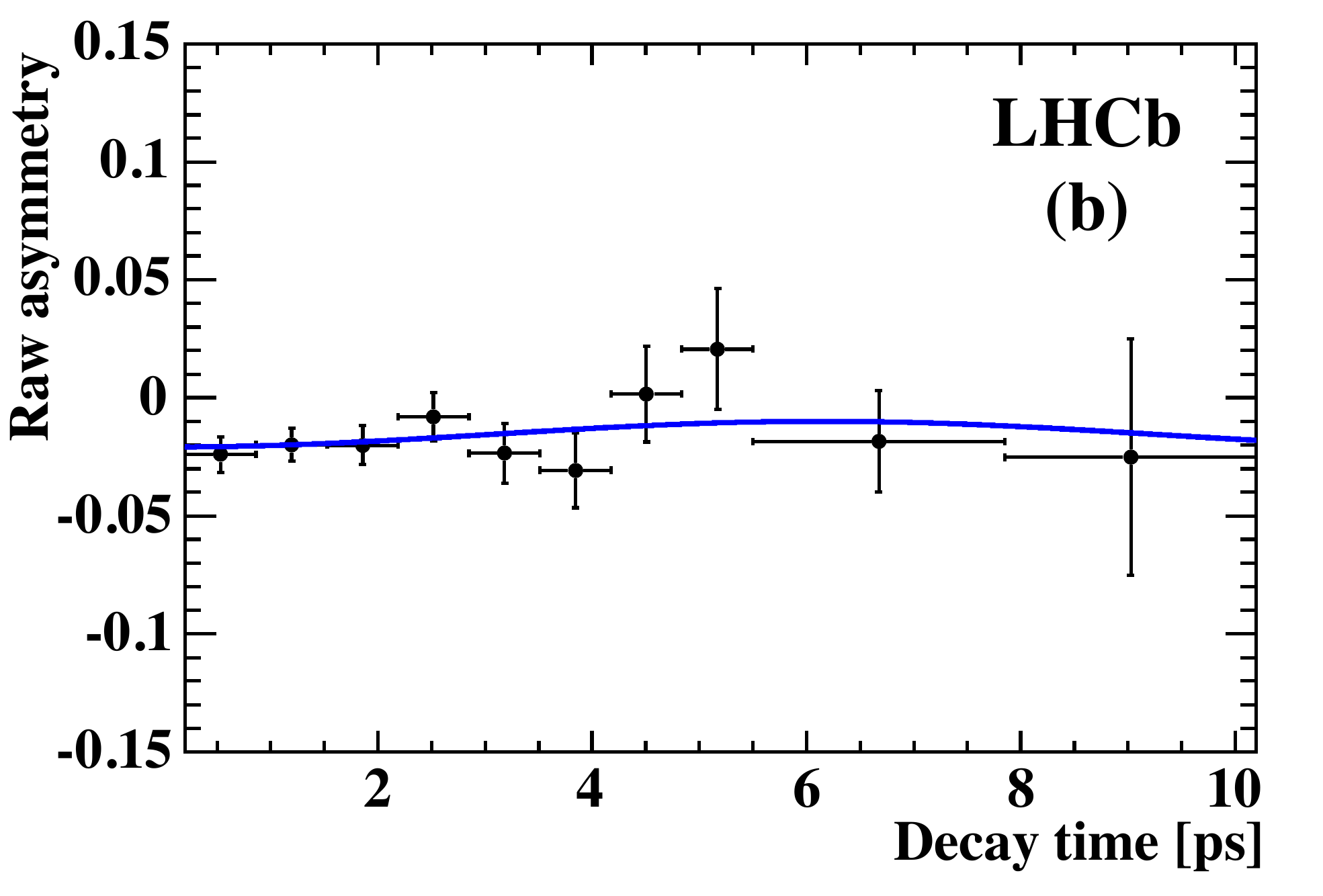}
   \includegraphics[width=0.48\linewidth]{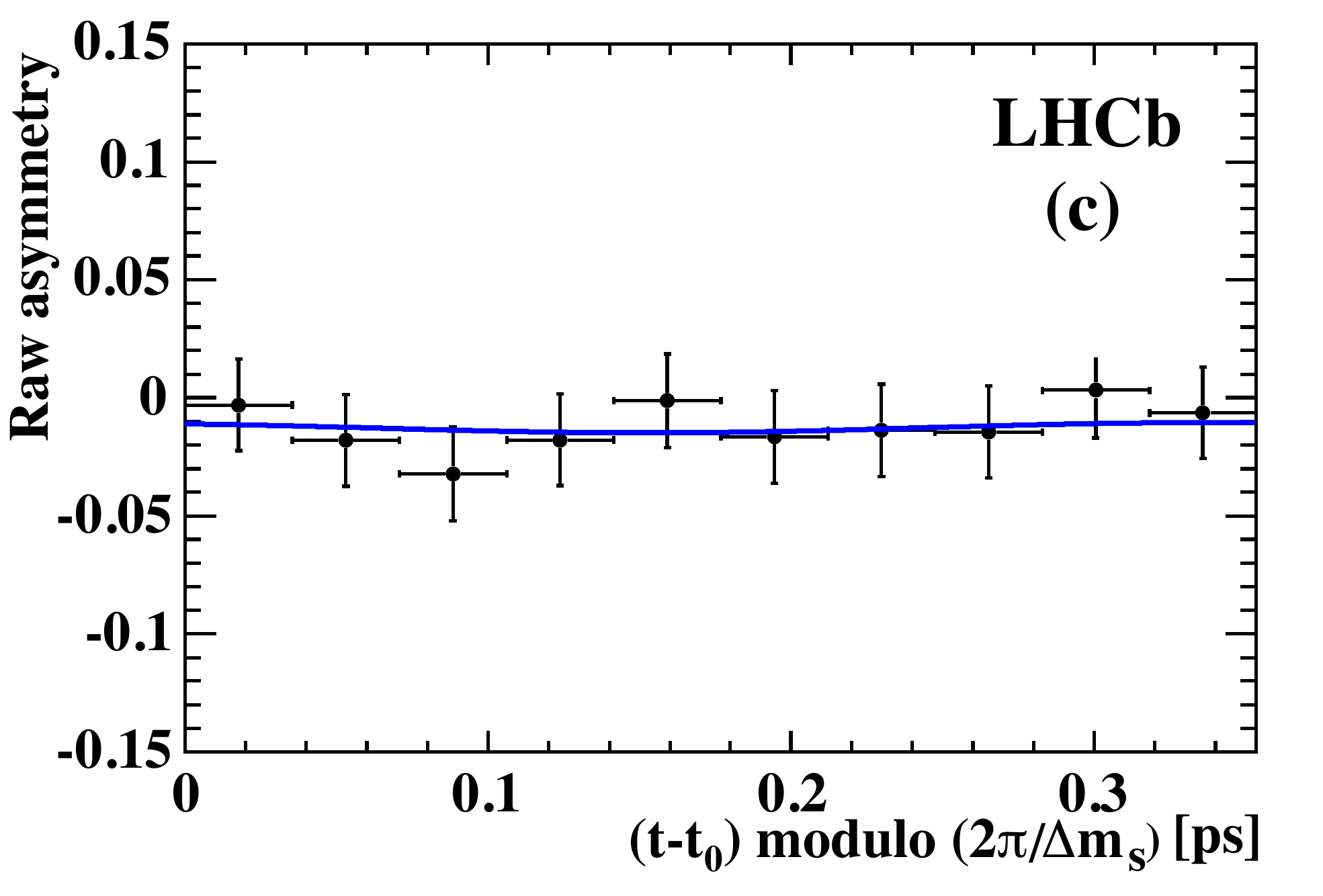}
  \end{center}
\caption{\small Time-dependent raw asymmetries for candidates in the (a) \BdToJPsiKst, (b) $\Bz \to \Dm \pip$ and (c) $\Bs \to \Dsm \pip$ signal mass regions with the results of the global fits overlaid. In (c) the asymmetry is obtained by folding the \Bsb and \Bs decay time distributions into one oscillation period, and
the offset $t_0 = 0.2$\ps corresponds to the selection requirement on the decay time.}
  \label{fig:resultBd2JpsiKstAsym}
\end{figure}

Figure~\ref{fig:resultBd2JpsiKst} shows the $\jpsi K^+\pi^-$, $K^+\pi^-\pi^-\pi^+$ and $K^+K^-\pi^-\pi^+$ invariant mass and decay time distributions, with the results of the global fits overlaid. Figure~\ref{fig:resultBd2JpsiKstAsym} shows the raw asymmetries, defined as the ratios between the difference and the sum of the overall decay time distributions, as a function of decay time for candidates in the signal mass region.  The signal yields, $A_{\rm P}$ values and detection asymmetries obtained from the global fits are reported in Table~\ref{tab:globalfit}. The $A_{\rm P}$ values obtained from the global fits are not well defined physical quantities, because efficiency corrections as a function of \pt and $\eta$ need to be applied. They are reported here for illustrative purposes only.  

\begin{table}[tb]
 \caption{\small Values of signal yields,  $A_{\rm P}$, $A_{f}$ and of the correlations $\rho(A_{\rm P},\,A_f)$ obtained from global fits. The smaller value of the correlation in the \Bs case is due to the much larger mixing frequency of \Bs mesons. }
 \begin{center}
   \begin{tabular}{cccc}
\hline
     Parameter & \BdToJPsiKst & $\Bz \to \Dm \pip$ & $\Bs \to \Dsm \pip$ \\
\hline
 $N^{\mathrm{sig}}$                      & $93\,627 \pm 360$        &  $76\,682 \pm 308 $                   & $16\,887 \pm 174$ \\
     $A_\mathrm{P}$                     & $-0.0116 \pm 0.0063$  & $-0.0058 \pm 0.0070$          & $-0.0032 \pm 0.0166$  \\	
     $A_f$                              & $-0.0086 \pm 0.0046$  & $-0.0151 \pm 0.0049$          & $-0.0110 \pm 0.0086$ \\
     $\rho(A_{\rm P},\,A_f)$      & $-0.65$ & $-0.64$ & $-0.01$ \\
\hline
 \label{tab:globalfit}
\end{tabular}
\end{center}
\end{table}

\begin{figure}[!t]
 \begin{center}
    \includegraphics[width=0.48\linewidth]{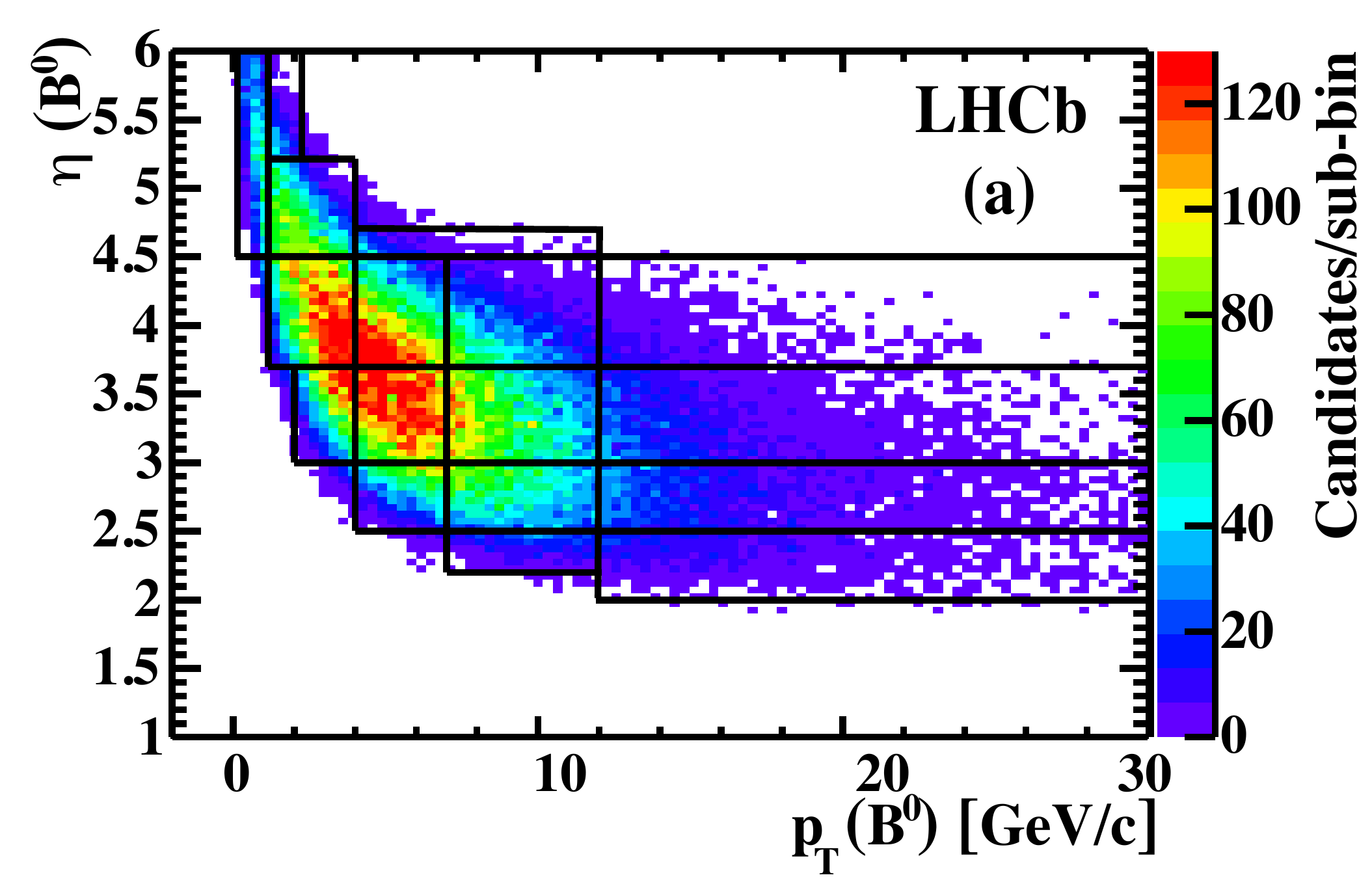} 
    \includegraphics[width=0.48\linewidth]{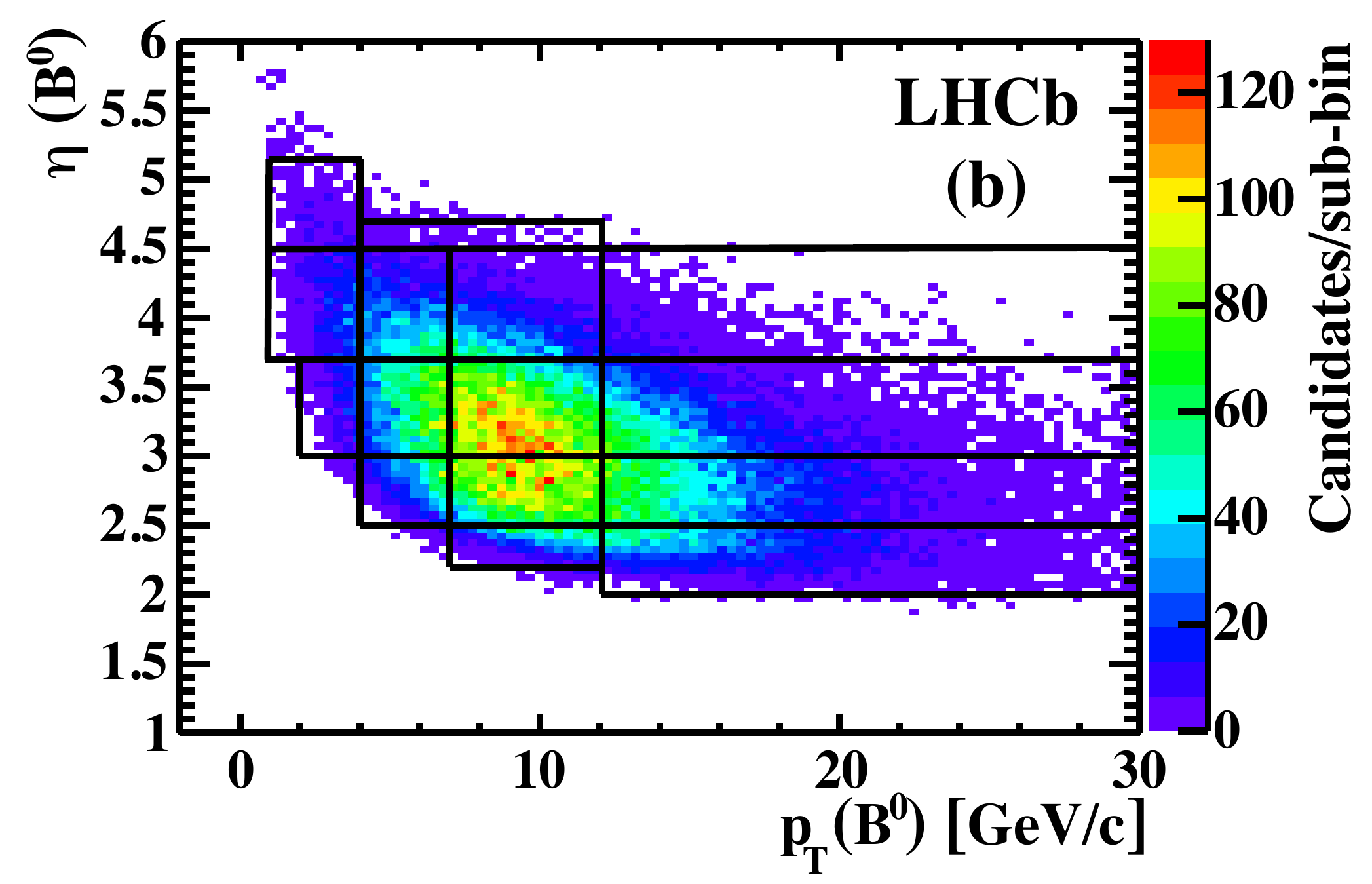} 
    \includegraphics[width=0.48\linewidth]{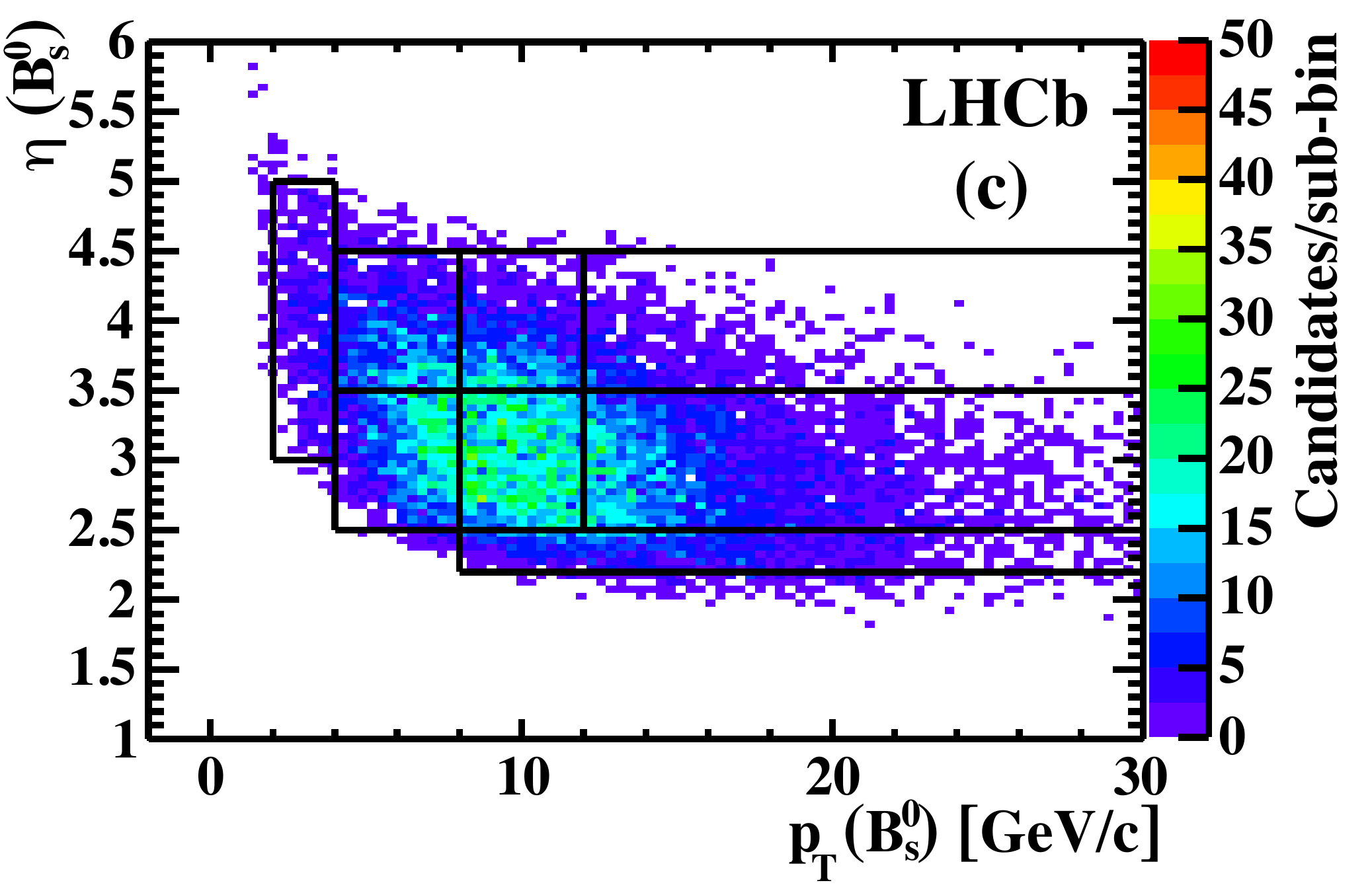}
    \vspace{-0.6cm} 
\end{center}
  \caption{\small Distributions of $p_\mathrm{T}$ and $\eta$, where the background components are subtracted using the \emph{sPlot} technique~\cite{Pivk:2004ty}, for  (a) \BdToJPsiKst , (b) $\Bz \to \Dm \pip$ and (c) $\Bs \to \Dsm \pip$ decays. The definition of the various kinematic bins is superimposed.} 
  \label{fig:Bbins}
\end{figure}

Figure~\ref{fig:Bbins} shows the two-dimensional distributions of ($p_\mathrm{T}$, $\eta$) for \BdToJPsiKst, $\Bz \to \Dm \pip$ and $\Bs \to \Dsm \pip$ decays. The background components are subtracted using the \emph{sPlot} technique~\cite{Pivk:2004ty} and the chosen definition of the various kinematic bins is overlaid. 
For the two \Bz decays we use a common set of bins, as reported in Table~\ref{tab:binsresultscombinedB0}, in order to allow a simple combination of the two independent  $A_\mathrm{P}$ measurements. In the case of the $\Bd \to \jpsi \Kstarz$, two additional bins at small \pt and large $\eta$ are also defined. An accurate knowledge of the decay time resolution is important for $\Bs \to \Dsm \pip$ decay, due to the fast oscillation of the \Bs meson. For this reason we determine the decay time resolution using the method previously described, applied to events belonging to each (\pt, $\eta$) bin, where a double Gaussian function with zero mean and values of the widths depending on the given bin is used.

\section{Systematic uncertainties}
\label{sec:systematic}
Several sources of systematic uncertainty that affect the determination of the production asymmetries are considered.
For the invariant mass model, the effects of the uncertainty on the shapes of all components (signals, combinatorial and partially reconstructed backgrounds) are investigated.
For the decay time model, systematic effects related to the decay time resolution and acceptance are studied. The effects of the uncertainties on the external inputs used in the fits, reported in Table~\ref{tab:input},
are evaluated by repeating the fits with each parameter varied by $\pm 1\sigma$. Alternative parameterizations of the background components are also considered.
To estimate the contribution of each single source, we repeat the fit for each (\pt, $\eta$) bin after having modified the baseline fit model. 
The shifts from the relevant baseline values are taken as the systematic uncertainties. 
To estimate a systematic uncertainty related to the parameterization of final state radiation effects on the signal mass distributions, the parameter $s$ of Eq.~\ref{eq:radcor2} is varied by  $\pm 1\sigma$  of the corresponding value obtained from fits to simulated events.
A systematic uncertainty related to the invariant mass resolution model is estimated by repeating the fit using a single Gaussian function.
The systematic uncertainty related to the parameterization of the mass shape for the combinatorial background 
is investigated by replacing the exponential function with a straight line. 
Concerning the partially reconstructed background, we assess a
systematic uncertainty by repeating the fits while excluding the low mass sideband, \emph{i.e.} applying the requirements $m>5.20$\gevcc
for the $\Bz \to \Dm \pip$ decays and $m>5.33$\gevcc for $\Bs \to \Dsm \pip$ decays. 
To estimate the uncertainty related to the parameterization of signal decay time acceptances, different acceptance functions are considered. 
Effects of inaccuracies in the knowledge of the decay time resolution are estimated by rescaling the widths of the baseline model to obtain an average resolution width differing by $\pm 8$\fs. Simulation studies also indicate that there is a small bias in the reconstructed decay time. The impact of such a bias is assessed by introducing a corresponding bias of $\pm 2$\fs in the decay time resolution model.

The determination of the systematic uncertainties related to the $|q/p|$ input value needs a special treatment, as $A_\mathrm{P}$ is correlated with $|q/p|$. For this reason, any variation of $|q/p|$ turns into the same shift of $A_\mathrm{P}$  in each of the kinematic bins.  Such a correlation is taken into account when averaging $A_\mathrm{P}(B^0)$ measurements from \BdToJPsiKst and $\Bd\to\Dm\pip$ decays, or when integrating over $p_\mathrm{T}$ and $\eta$. The values of the systematic uncertainties related to the knowledge of  $|q/p|$ are 0.0013 in the case of $A_{\rm P}(\Bz)$ and 0.0030 in the case of $A_{\rm P}(\Bs)$.
The dominant systematic uncertainties for the \BdToJPsiKst decay are related to the signal mass shape and to $|q/p|$.
For the $\Bz \to \Dm \pip$ decay, the most relevant systematic uncertainties are related to the signal mass shape and to the partially reconstructed background. Systematic uncertainties associated with the decay time resolution and $\Delta m_s$ are the main sources for the $\Bs \to \Dsm \pip$ decay. 

\section{Results}

The values of $A_\mathrm{P}(B^0)$ are determined independently for  \BdToJPsiKst and  $\Bz \to \Dm \pip$ decays in each kinematic bin and then averaged.
Table~\ref{tab:binsresultscombinedB0} reports the final results. The overall bin-by-bin agreement between the two sets of independent $A_\mathrm{P}(B^0)$ measurements is evaluated by means of a $\chi^2$ test, with a $\chi^2=7$ for $14$ degrees of freedom. The values of $A_\mathrm{P}(B^0_s)$ determined from the $\Bs \to \Dsm \pip$ fits are reported in Table~\ref{tab:binsresultsBs}.

\begin{table}[t]
 \caption{\small Combined values of $A_\mathrm{P}(\Bz)$ from $B^0\rightarrow J/\psi K^{*0}$ and $\Bz \to \Dm \pip$ decays, corresponding to the various kinematic bins. The first uncertainties are statistical and the second systematic. For completeness, the values obtained either from $B^0\rightarrow J/\psi K^{*0}$ or $\Bz \to \Dm \pip$ decays are also reported in the last two columns, with statistical uncertainties only. The values of the last two bins are obtained from $B^0\rightarrow J/\psi K^{*0}$ decays alone.}  
\vspace{-0.4cm}
 \begin{center}
\resizebox{1\textwidth}{!}{
   \begin{tabular}{ccccc} 
\hline
\pt (\gevc) & $\eta$  & $A_\mathrm{P}(\Bz)$ & $A_\mathrm{P}(\Bz \to J/\psi K^{*0})$ & $A_\mathrm{P}(\Bz \to \Dm \pip)$\\ 
\hline
$(\phantom{1}1.0,\phantom{1}4.0)$&$(4.5,5.2)$ & $\phantom{-}0.0016 \pm 0.0253 \pm 0.0016$ & $\phantom{-}0.0037 \pm 0.0260$ & $-0.0331 \pm 0.1044$ \\
$(\phantom{1}1.0,\phantom{1}4.0)$&$(3.7,4.5)$ & $-0.0158 \pm 0.0162 \pm 0.0015$ & $-0.0161 \pm 0.0170$ & $-0.0130 \pm 0.0519$ \\
$(\phantom{1}2.0,\phantom{1}4.0)$&$(3.0,3.7)$ & $\phantom{-}0.0055 \pm 0.0254 \pm 0.0016$ & $\phantom{-}0.0078 \pm 0.0271 $ & $-0.0114 \pm 0.0738$ \\
$(\phantom{1}4.0,12.0)$&$(4.5,4.7)$ & $\phantom{-}0.0160 \pm 0.0736 \pm 0.0067$ & $-0.0489 \pm 0.0840$ & $\phantom{-}0.2353 \pm 0.1529 $ \\
$(\phantom{1}4.0,\phantom{1}7.0)$&$(3.7,4.5)$ & $-0.0189 \pm 0.0158 \pm 0.0032$ & $-0.0221 \pm 0.0184 $ & $-0.0099 \pm 0.0310$ \\
$(\phantom{1}4.0,\phantom{1}7.0)$&$(3.0,3.7)$ & $-0.0311 \pm 0.0132 \pm 0.0014$ & $-0.0342 \pm 0.0160$ & $-0.0245 \pm 0.0232$ \\
$(\phantom{1}4.0,\phantom{1}7.0)$&$(2.5,3.0)$ & $\phantom{-}0.0556 \pm 0.0254 \pm 0.0020$ & $\phantom{-}0.0703 \pm 0.0324 $ & $\phantom{-}0.0321 \pm 0.0408$ \\
$(\phantom{1}7.0,12.0)$&$(3.7,4.5)$ & $-0.0145 \pm 0.0205 \pm 0.0027$ & $-0.0364 \pm 0.0269 $ & $\phantom{-}0.0161 \pm 0.0316$ \\
$(\phantom{1}7.0,12.0)$&$(3.0,3.7)$ & $-0.0142 \pm 0.0111 \pm 0.0015$ & $-0.0067 \pm 0.0173$ & $-0.0196 \pm 0.0146$ \\
$(\phantom{1}7.0,12.0)$&$(2.5,3.0)$ & $-0.0236 \pm 0.0138 \pm 0.0014$ & $-0.0341 \pm 0.0228$ & $-0.0175 \pm 0.0173$ \\
$(\phantom{1}7.0,12.0)$&$(2.2,2.5)$ & $-0.0190 \pm 0.0348 \pm 0.0034$ & $-0.0397 \pm 0.0623$ & $-0.0096 \pm 0.0420$ \\
$(12.0,30.0)$&$(3.7,4.5)$ & $-0.0550 \pm 0.0473 \pm 0.0020$ & $-0.0195 \pm 0.0649$ & $-0.0951 \pm 0.0690 $ \\
$(12.0,30.0)$&$(3.0,3.7)$ & $\phantom{-}0.0067 \pm 0.0180 \pm 0.0021$ & $-0.0193 \pm 0.0311$ & $\phantom{-}0.0199 \pm 0.0220$ \\
$(12.0,30.0)$&$(2.5,3.0)$ & $\phantom{-}0.0177 \pm 0.0162 \pm 0.0023$ & $\phantom{-}0.0295 \pm 0.0314$ & $\phantom{-}0.0134 \pm 0.0190$ \\
$(12.0,30.0)$&$(2.0,2.5)$ & $-0.0018 \pm 0.0236 \pm 0.0020$ & $\phantom{-}0.0031 \pm 0.0485$ & $-0.0033 \pm 0.0270$ \\
$( \phantom{1}0.2,\phantom{1}1.0)$&$(4.5,6.0)$& $-0.0391 \pm 0.0501 \pm 0.0016$ & $-0.0391 \pm 0.0501$ & $\phantom{-}-$ \\
$(\phantom{1}1.0,\phantom{1}2.2)$&$(5.2,6.0)$& $\phantom{-}0.0523 \pm 0.0684 \pm 0.0025$ & $\phantom{-}0.0523 \pm 0.0684$ & $\phantom{-}-$ \\
\hline
\end{tabular}
}
\vspace{-0.4cm}
\end{center}
 \label{tab:binsresultscombinedB0}
\end{table}

\begin{table}[t]
\caption{\small Values of $A_\mathrm{P}(\Bs)$ from $\Bs \to \Dsm \pip$ decays, corresponding to the various kinematic bins. The first uncertainties are statistical and the second systematic.}  
\vspace{-0.4cm}
 \begin{center}
   \begin{tabular}{ccc}
\hline
\pt (\gevc) & $\eta$  & $A_{\rm P}(\Bs)$  \\ 
\hline
$(\phantom{1}2,\phantom{1}4)$ & $(3.0,5.0)$ & $  -0.1475 \pm 0.0895 \pm 0.0192 $ \\
$(\phantom{1}4,\phantom{1}8)$ & $(3.5,4.5)$ & $  -0.0471 \pm 0.0513 \pm 0.0112 $ \\
$(\phantom{1}4,\phantom{1}8)$ & $(2.5,3.5)$ &$  \phantom{-}0.0376 \pm 0.0467 \pm 0.0083$  \\
$(\phantom{1}8,12)$ & $(3.5,4.5)$ & $  \phantom{-}0.0582 \pm 0.0537 \pm 0.0053 $   \\
$(\phantom{1}8,12)$ & $(2.5,3.5)$ &$  \phantom{-}0.0370 \pm 0.0332 \pm 0.0051$  \\
$(12,30)$ &$(3.5,4.5)$ & $  -0.0339 \pm 0.0750  \pm 0.0095$ \\
$(12,30)$ & $(2.5,3.5)$ &$  -0.0333 \pm 0.0309 \pm 0.0040$ \\
$(\phantom{1}8,30)$ & $(2.2,2.5)$ &$  -0.0351 \pm 0.0485 \pm 0.0059$ \\
\hline
\end{tabular}\end{center}
 \label{tab:binsresultsBs}
\end{table}

The integration over \pt and $\eta$ of the bin-by-bin $A_\mathrm{P}$ values is performed within the ranges $4 < p_\mathrm{T} < 30$\gevc and $2.5 < \eta < 4.5$. The integrated value of $A_\mathrm{P}$ is given by 
\begin{equation}
\label{eq:production}
A_\mathrm{P} = \frac{\sum_i\frac{N_i}{\varepsilon_i} A_{\mathrm{P},i}}{\sum_i\frac{N_i}{\varepsilon_i}},
\end{equation}
where the index $i$ runs over the bins, $N_i$ is the number of
signal events and $\varepsilon_i$ is the efficiency, defined as the number of selected events divided by the number of produced events in the \emph{i}-th bin.
The signal yield in each bin can be expressed as
\begin{equation}
\label{eq:yield_i}
N_i = \mathcal{L} \cdot \sigma_{b\bar{b}} \cdot 2 \cdot f_{d(s)} \cdot \mathcal{B} \cdot  f_i \cdot \varepsilon_i,
\end{equation}
where $\mathcal{L}$ is the integrated luminosity, $\sigma_{b\bar{b}}$ is the $b\bar{b}$ cross section, $f_{d(s)}$ is the $B^0_{(s)}$ hadronization fraction, $f_i$ is the fraction of $B$ mesons produced in the \emph{i}-th bin and $\mathcal{B}$ is the branching fraction of the $B$ decay.
By substituting $N_i /  \varepsilon_i$ from Eq.~\ref{eq:yield_i} into  Eq.~\ref{eq:production}, the integrated value of $A_\mathrm{P}$ becomes
\begin{equation}
\label{eq:production_weights}
A_\mathrm{P} = \sum_i \omega_i A_{\mathrm{P},i}\,,
\end{equation}
where $\omega_i = f_i / \sum_i f_i$.
The values of $\omega_i$ are determined using simulated events. The difference between the values of $\omega_i$ predicted by \pythia for $B^0$ and $B^0_s$ mesons is found to be negligible, if the same bins in \pt and $\eta$ would be used.
\begin{table}[t]
 \caption{\small Values of $\omega_i$ determined from simulation and  $\omega_i^{\rm data}$ extracted from data using $B^0\rightarrow J/\psi K^{*0}$ decays in two different binning schemes. The $\omega_i$ values and the difference between $\omega_i$ and $\omega_i^{\rm data}$ values are used to determine the integrated results and to evaluate the related systematic uncertainties, respectively.
} 
\vspace{-0.2cm} 
 \begin{center}
   \begin{tabular}{cccc}
\hline
 \pt (\gevc) & $\eta$  & $\omega_i$ & $\omega_i^{\rm data}$   \\
\hline
$(\phantom{1}4, \phantom{1}7)$ & $(3.7, 4.5)$ & $0.1698 \pm 0.0008$  & $0.1946 \pm 0.0025$ \\
$(\phantom{1}4, \phantom{1}7)$ & $(3.0, 3.7)$ & $0.2432 \pm 0.0009$  & $ 0.2396 \pm 0.0036$\\
$(\phantom{1}4, \phantom{1}7)$ & $(2.5, 3.0)$ & $0.2222 \pm 0.0009$   & $0.1976 \pm 0.0051$\\ 
$(\phantom{1}7, 12)$ & $(3.7, 4.5)$ & $0.0662 \pm 0.0006$   & $ 0.0789 \pm 0.0016$ \\ 
$(\phantom{1}7, 12)$ & $(3.0, 3.7)$& $0.1129 \pm 0.0007$   & $0.1129 \pm 0.0045$\\ 	
$(\phantom{1}7, 12)$ & $(2.5, 3.0)$ & $0.1150 \pm 0.0007$   & $0.1002 \pm 0.0019$\\ 
$(12, 30)$ & $(3.7, 4.5)$ & $0.0113 \pm 0.0003$   & $0.0160 \pm 0.0007$\\ 
$(12 ,30)$ & $(3.0, 3.7)$ & $0.0276 \pm 0.0004$  & $0.0307 \pm 0.0028$\\
$(12, 30)$ & $(2.5, 3.0)$ & $0.0318 \pm 0.0004$  & $0.0296 \pm 0.0025$\\				
\hline
$(\phantom{1}4, \phantom{1}8)$ & $(3.5, 4.5)$ & $0.2667 \pm 0.0009$ & $0.3064 \pm 0.0020$ \\
$(\phantom{1}4, \phantom{1}8)$ & $(2.5, 3.5)$ & $0.4766 \pm 0.0009$ & $0.4644 \pm 0.0030$ \\
$(\phantom{1}8, 12)$ & $(3.5, 4.5)$ & $0.0564 \pm 0.0005$ & $0.0640 \pm 0.0015$ \\
$(\phantom{1}8, 12)$ & $(2.5, 3.5)$ & $0.1295 \pm 0.0008$ & $0.0873 \pm 0.0019$ \\
$(12, 30)$ & $(3.5, 4.5)$ & $0.0175 \pm 0.0003$ & $0.0238 \pm 0.0008$ \\
$(12, 30)$ & $(2.5, 3.5)$ & $0.0532 \pm 0.0005$ & $0.0541 \pm 0.0027$ \\
\hline
\end{tabular}\end{center}
 \label{tab:w_i_B0}
\end{table}
These values are also extracted from data using \BdToJPsiKst decays. In this case $\omega_i^{\rm data}$ is measured as
\begin{equation}
\omega_i^\mathrm{data} = \frac{N_i}{\varepsilon^{\rm rec}_i}\,\,/\,\,\sum_j{\frac{N_j}{\varepsilon^{\rm rec}_j}},
\end{equation} 
where $N_i$ is the yield in the \emph{i}-th bin and $\varepsilon_i^{\rm rec}$ is total reconstruction efficiency.
The values of $\varepsilon^{\rm rec}_{i}$  are determined using both simulated events and data control samples. The values of $\omega_i$ and  $\omega_i^\mathrm{data}$, summarized in Table~\ref{tab:w_i_B0}, exhibit systematic differences at the 10\% level. The difference in the central value between $A_\mathrm{P}(\BdToJPsiKst)$ calculated using either  $\omega_i$ or $\omega_i^\mathrm{data}$ is found to be 0.0024 using the $\Bz$ binning scheme, and 0.0034 using the $\Bs$ binning scheme. These values are assigned as systematic uncertainties for $A_\mathrm{P}(\Bz)$ and $A_\mathrm{P}(\Bs)$. Table~\ref{tab:totsyst} summarizes the systematic uncertainties associated with the integrated measurements. In the first row, the combined systematic uncertainties estimated in each bin, as described in the previous section, are reported.

\begin{table}[t]
 \caption{\small Absolute values of systematic uncertainties. The total systematic uncertainties are obtained by summing the individual contributions in quadrature.}  
\vspace{-0.4cm}
 \begin{center}
   \begin{tabular}{ccc} 
\hline
Source & \multicolumn{2}{c}{Uncertainty}\\
 & $A_\mathrm{P}(\Bz)$ & $A_\mathrm{P}(\Bs)$\\
\hline
Combined systematic uncertainties from bin studies & 0.0004 & 0.0048 \\
Uncertainty on $|q/p|$ & 0.0013 & 0.0030\\
Difference between $\omega_i$ and $\omega^{\rm data}_i$  & 0.0024 & 0.0034 \\
\hline
Total & 0.0028 & 0.0066 \\
\hline
\end{tabular}
\vspace{-0.4cm}
\end{center}
 \label{tab:totsyst}
\end{table}

\begin{figure}[h!]
 \begin{center}
    \includegraphics[width=0.42\linewidth]{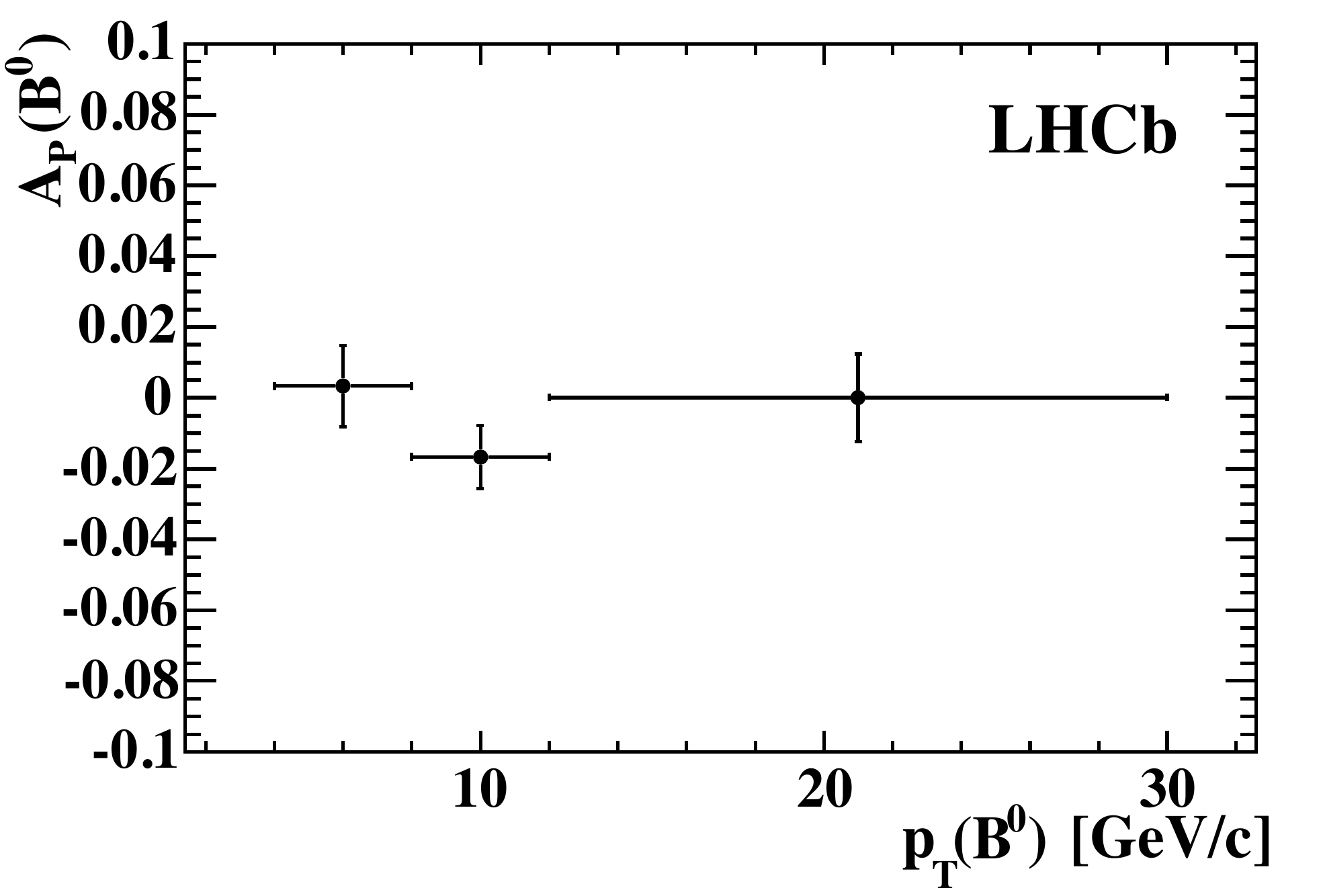}
    \includegraphics[width=0.42\linewidth]{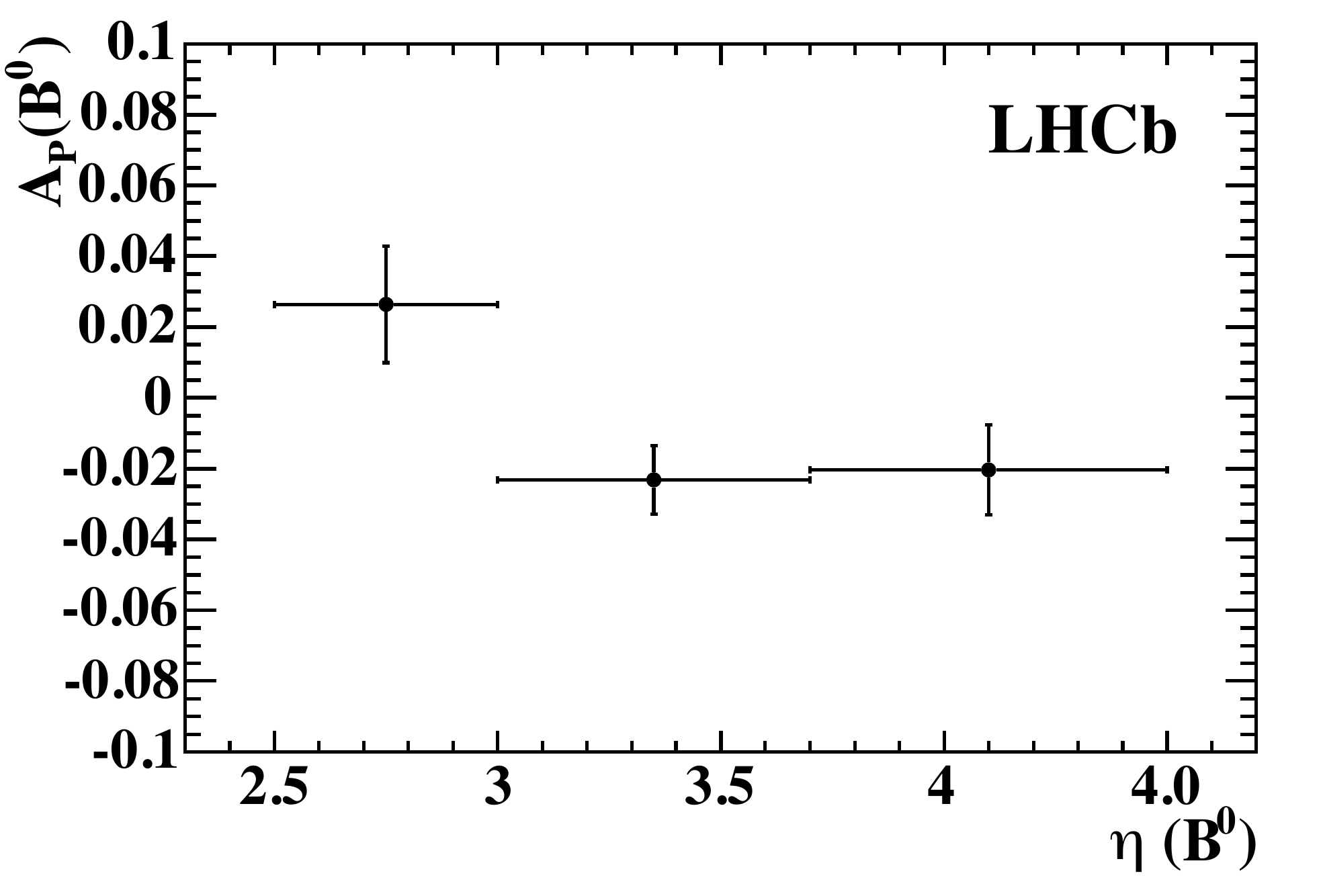}
    \includegraphics[width=0.42\linewidth]{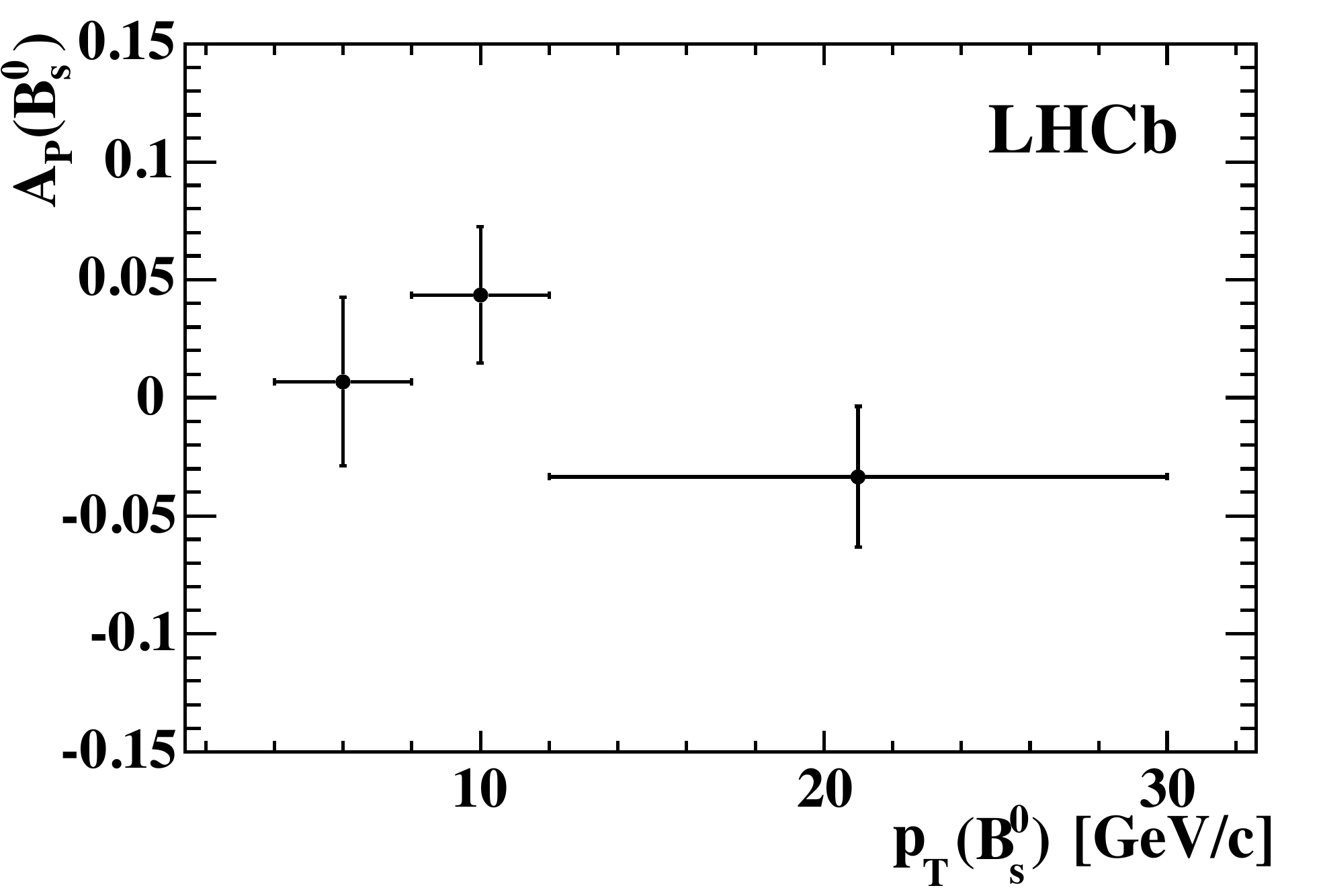}
    \includegraphics[width=0.42\linewidth]{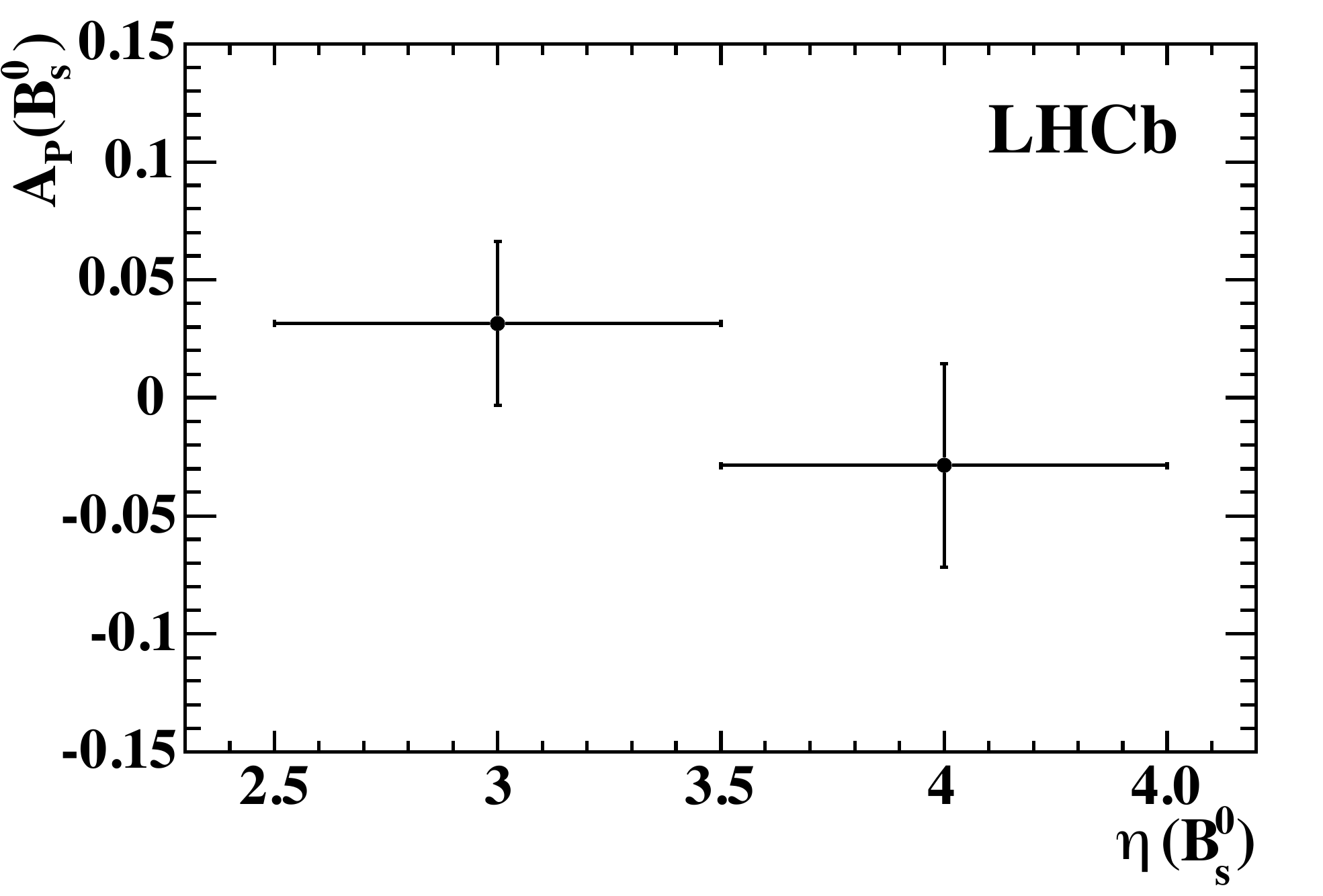}
\vspace{-0.5cm}
\end{center}
  \caption{\small Dependence of (top) $A_\mathrm{P}(B^0)$  and (bottom) $A_\mathrm{P}(B^0_s)$ on (left) \pt and (right) $\eta$. The error bars include both statistical and systematic uncertainties.}
  \label{fig:B0dep}
\end{figure}

Using Eq.~\ref{eq:production_weights}, the integrated measurements of $A_\mathrm{P}(B^0)$ for \BdToJPsiKst and $\Bz \to \Dm \pip$ decays are found to be 
\begin{eqnarray*}
A_\mathrm{P}(\BdToJPsiKst) = & -0.0033 \pm 0.0096 \,\mathrm{(stat)} \pm 0.0028 \, \mathrm{(syst)} , \\
A_\mathrm{P}(\Bz \to \Dm \pip) = & -0.0038 \pm 0.0124 \,\mathrm{(stat)} \pm 0.0029 \, \mathrm{(syst)},
\end{eqnarray*}
which lead to the average 
\begin{equation}
A_\mathrm{P}(\Bz)= -0.0035 \pm 0.0076 \,\mathrm{(stat)} \pm 0.0028 \,\mathrm{(syst)}.\nonumber
\end{equation}
The integrated value of $A_\mathrm{P}(\Bs)$ is 
\begin{equation}
A_\mathrm{P}(\Bs)=\phantom{-}0.0109 \pm 0.0261 \,\mathrm{(stat)} \pm 0.0066 \,\mathrm{(syst)}.\nonumber
\end{equation}
Finally, the dependencies of $A_\mathrm{P}(B^0)$ and $A_\mathrm{P}(B^0_s)$ on \pt, obtained by integrating over $\eta$, and on $\eta$, obtained by integrating over \pt, are shown in Fig.~\ref{fig:B0dep}. The corresponding numerical values are reported in Tables~\ref{tab:integral_over_pt_eta} and~\ref{tab:integral_over_pt_eta_bs}.

\begin{table}[h]
\caption{\small Values of the production asymmetry $A_\mathrm{P}(B^0)$ in bins of \pt and $\eta$ from $\Bz\to J/\psi\Kstarz$ and $\Bd\to\Dm\pip$ decays. The first uncertainties are statistical and the second systematic.}
\begin{center}
   \begin{tabular}{ccc}
\hline
Variable & Bin &  $A_\mathrm{P}(\Bz)$\\
\hline
\multirow{3}{*}{\pt (\!\gevc)} & $(\phantom{1}4,\phantom{1}7)$      & $\phantom{-}0.0033 \pm 0.0111 \pm 0.0028$  \\
& $(\phantom{1}7,12)$          &  $-0.0167 \pm 0.0084 \pm 0.0028$ \\
& $(12,30)$       & $\phantom{-}0.0001 \pm 0.0130 \pm 0.0029$ \\
\hline
\multirow{3}{*}{$\eta$} & $(2.5,3.0)$& $\phantom{-}0.0264 \pm 0.0161 \pm 0.0030$\\
 & $(3.0,3.7)$     & $-0.0232 \pm 0.0093 \pm 0.0028$ \\
& $(3.7,4.5)$   & $-0.0203 \pm 0.0125 \pm 0.0021$ \\
\hline
\end{tabular}
\end{center}
 \label{tab:integral_over_pt_eta}
\end{table}

\begin{table}[h]
 \caption{\small Values of the production asymmetry $A_\mathrm{P}(B^0_s)$ in bins of \pt and $\eta$ from $\Bs\to\Dsm\pip$ decays. The first uncertainties are statistical and the second systematic.}
 \begin{center}
   \begin{tabular}{ccc}
\hline
Variable & Bin & $A_{\rm P} (B^0_s)$ \\
\hline
\multirow{3}{*}{\pt (\!\gevc)}  & $(\phantom{1}4,\phantom{1}8)$    & $\phantom{-}0.0069 \pm 0.0351 \pm 0.0067$ \\
& $(\phantom{1}8,12)$   & $\phantom{-}0.0435 \pm 0.0283 \pm 0.0039 $ \\
& $(12,30)$  & $-0.0334 \pm 0.0296 \pm 0.0038$  \\
\hline
\multirow{2}{*}{$\eta$} & (2.5, 3.5) & $\phantom{-}0.0315 \pm 0.0342 \pm 0.0060 $ \\
& (3.5, 4.5) & $-0.0286 \pm 0.0412 \pm 0.0088 $\\
\hline
\end{tabular}\end{center}
 \label{tab:integral_over_pt_eta_bs}
\end{table}


\section{Conclusions}

The production asymmetries of \Bz and \Bs mesons have been measured in $pp$ collisions at $\sqrt{s}=7$\tev within the acceptance of the LHCb detector, using a data
sample corresponding to an integrated luminosity of 1.0\invfb. The measurements have been performed in bins of \pt and $\eta$, and provide constraints that can be used to test different models of $B$-meson production.
Furthermore, once integrated using appropriate weights for any reconstructed $B^0_{(s)}$ decay mode, they can be used to derive effective production asymmetries, as inputs for \CP violation measurements with the LHCb detector.

The values of the production asymmetries integrated in the ranges $4 < p_\mathrm{T} < 30$\gevc and $2.5<\eta<4.5$ have been determined to be
\begin{equation}
A_\mathrm{P}(\Bz)= (-0.35 \pm 0.76  \pm 0.28)\%,\nonumber
\end{equation}
\begin{equation}
A_\mathrm{P}(\Bs)= (\phantom{-}1.09 \pm 2.61  \pm 0.66 )\%,\nonumber
\end{equation}
where the first uncertainties are statistical and the second systematic. No clear evidence of dependences on the values of \pt and $\eta$ has been observed.

\section{Acknowledgments}

\noindent We express our gratitude to our colleagues in the CERN
accelerator departments for the excellent performance of the LHC. We
thank the technical and administrative staff at the LHCb
institutes. We acknowledge support from CERN and from the national
agencies: CAPES, CNPq, FAPERJ and FINEP (Brazil); NSFC (China);
CNRS/IN2P3 (France); BMBF, DFG, HGF and MPG (Germany); SFI (Ireland); INFN (Italy); 
FOM and NWO (The Netherlands); MNiSW and NCN (Poland); MEN/IFA (Romania); 
MinES and FANO (Russia); MinECo (Spain); SNSF and SER (Switzerland); 
NASU (Ukraine); STFC (United Kingdom); NSF (USA).
The Tier1 computing centres are supported by IN2P3 (France), KIT and BMBF 
(Germany), INFN (Italy), NWO and SURF (The Netherlands), PIC (Spain), GridPP 
(United Kingdom).
We are indebted to the communities behind the multiple open 
source software packages on which we depend. We are also thankful for the 
computing resources and the access to software R\&D tools provided by Yandex LLC (Russia).
Individual groups or members have received support from 
EPLANET, Marie Sk\l{}odowska-Curie Actions and ERC (European Union), 
Conseil g\'{e}n\'{e}ral de Haute-Savoie, Labex ENIGMASS and OCEVU, 
R\'{e}gion Auvergne (France), RFBR (Russia), XuntaGal and GENCAT (Spain), Royal Society and Royal
Commission for the Exhibition of 1851 (United Kingdom).


\bibliographystyle{LHCb}
\bibliography{main}

\clearpage

\centerline{\large\bf LHCb collaboration}
\begin{flushleft}
\small
R.~Aaij$^{41}$, 
B.~Adeva$^{37}$, 
M.~Adinolfi$^{46}$, 
A.~Affolder$^{52}$, 
Z.~Ajaltouni$^{5}$, 
S.~Akar$^{6}$, 
J.~Albrecht$^{9}$, 
F.~Alessio$^{38}$, 
M.~Alexander$^{51}$, 
S.~Ali$^{41}$
% ...

{\footnotesize \it
% ...
$ ^{5}$Clermont Universit\'{e}, Universit\'{e} Blaise Pascal, CNRS/IN2P3, LPC, Clermont-Ferrand, France\\
$ ^{6}$CPPM, Aix-Marseille Universit\'{e}, CNRS/IN2P3, Marseille, France\\
% ...
$ ^{9}$Fakult\"{a}t Physik, Technische Universit\"{a}t Dortmund, Dortmund, Germany\\
% ...
$ ^{37}$Universidad de Santiago de Compostela, Santiago de Compostela, Spain\\
$ ^{38}$European Organization for Nuclear Research (CERN), Geneva, Switzerland\\
% ...
$ ^{41}$Nikhef National Institute for Subatomic Physics, Amsterdam, The Netherlands\\
% ...
$ ^{46}$H.H. Wills Physics Laboratory, University of Bristol, Bristol, United Kingdom\\
% ...
$ ^{51}$School of Physics and Astronomy, University of Glasgow, Glasgow, United Kingdom\\
$ ^{52}$Oliver Lodge Laboratory, University of Liverpool, Liverpool, United Kingdom\\
% ...
}
\end{flushleft}

\end{document}